\documentstyle[12pt]{article}
\begin{document}
\newcommand{\newc}{\newcommand}

\newc{\be}{\begin{equation}}
\newc{\ee}{\end{equation}}
\newc{\ba}{\begin{eqnarray}}
\newc{\ea}{\end{eqnarray}}
\newc{\ie}{{\it i.e. }}
\newc{\eg}{{\it eg }}
\newc{\etc}{{\it etc.}}
\newc{\etal}{{\it et al.}}

\newc{\ra}{\rightarrow}
\newc{\lra}{\leftrightarrow}
\newc{\no}{Nielsen-Olesen }
\newc{\lsim}{\buildrel{<}\over{\sim}}
\newc{\gsim}{\buildrel{>}\over{\sim}}

\begin{titlepage}
\begin{center}
October 1994\hfill
MIT-CTP-2375\\
            \hfill
BROWN-HET-971
\vskip 1in

{\large \bf
Cosmic String Theory: The Current Status
}

\vskip .4in
{\large Leandros Perivolaropoulos}\footnote{E-mail address:
leandros@mitlns.mit.edu},\footnote{Also,
Department of Physics, Brown University, Providence, R.I. 02912.}\\[.15in]

{\em Center for Theoretical Physics\\
Laboratory for Nuclear Science\\
Massachussetts Institute of Technology\\
Cambridge, Mass. 02139, USA.}
\end{center}
\vskip .2in
\begin{abstract}
\noindent
This is a pedagogical introduction to the basics of the cosmic string theory
and also a review of the recent progress made with respect to the macrophysical
predictions of the theory. Topics covered include, string formation and
evolution, large scale structure formation, generation of peculiar velocity
flows, cosmic microwave background (CMB) fluctuations, lensing, gravitational
waves and constraints from the msec pulsar. Particular emphasis is placed on
the
signatures predicted on the CMB and the corresponding non-gaussian features.
\end{abstract}
\vspace{1.5cm}
\begin{center}
{\large \it Based on four lectures presented at the Trieste 1994 Summer School
for Particle Physics and Cosmology. To appear in the School Proceedings.}
\end{center}
\end{titlepage}

\newpage
\tableofcontents

\section{\bf Introduction-Cosmology Basics}

The Big Bang theory\cite{w72} is the best theory we have for describing the
universe. It is a
particularly simple and highly successful theory. It makes three basic
observationally consistent
 assumptions and derives from
them some highly non-trivial predictions which have been verified by
observations.  The main
assumptions are the following:
\begin{itemize}
\item
The universe is a homogeneous and isotropic thermal bath of the known particles
(cosmological
principle \cite{m33}).
\item
General Relativity is the correct theory to describe physics on cosmological
scales.
\item
The energy momentum tensor of the universe is well approximated by that of a
perfect
fluid \ie
\be
T_{ij}=diag [\rho, -p, -p, -p]
\ee
\end{itemize}
The first two assumptions imply that the universe can be described by the
Robertson-Walker
metric of the form
\be
ds^2=dt^2 - a(t)^2[{{dr^2}\over {1-kr^2}}+r^2(d\theta^2-\sin^2\theta d\phi^2)]
\ee
where $a(t)$ is known as {\it the scale factor of the universe} and $k$ can be
normalized
to the values $-1,0,1$ for an open (negative curvature), spatially flat and
closed (positive
curvature) universe respectively.

The third assumption can be used along with the Einstein equations to derive
the Friedman
equations that determine the dynamics of the scale factor $a(t)$. These may be
written as
\ba
 ({{\dot a}\over a})^2 & - &{k\over {a^2}}  =  {{8\pi G}\over 3} \rho  \\
 {{\ddot a}\over a}  &=&  {{4\pi G}\over 3} (\rho + 3p)
\ea

These are two equations with three unknown functions $(a(t),\rho(t),p(t))$ and
therefore the
equation of state $p=p(\rho)$ is also needed in order to obtain a solution. For
example, for
a flat $(k=0)$ radiation dominated universe $(p=\rho /3)$, it is easy to show
that
$a(t)\sim t^{1/2}$ while for a matter dominated universe $(p=0)$ we have
$a(t)\sim t^{2/3}$.

Observations indicate that at the present time the matter density dominates
over radiation
density in the universe $(\rho_{mat} (t_0)\gg \rho_{rad} (t_0))$ and therefore
we live in a
matter dominated universe. However, this was not the case at early times. The
Friedman
equations may be used to show that
\be
{d\over {da}} (\rho a^3) =- 3a^2 p
\ee
which implies
\ba
\rho_{mat} (t) &\sim & a(t)^{-3}\\
\rho_{rad} (t) &\sim &a(t)^{-4}
\ea
and therefore there is a time $t_{eq}<t_0$ of {\it equal matter and radiation}
such that
\be
t<t_{eq} \Rightarrow \rho_{mat} (t) < \rho_{rad} (t)
\ee

Using Eq. (7) and the well known result from statistical mechanics that
$\rho_{rad}\sim
T^4$ we obtain that the temperature of radiation scales as
\be
T(t)\sim a(t)^{-1}
\ee
in a matter or radiation dominated universe.
Cosmic Microwave Background (CMB) measurements\cite{m90} have shown that
$T(t_0)=2.736\pm 0.017$ and
therefore the temperature at earlier times is predicted to be
\be
T(t)=2.7^0 K {{a(t_0)}\over {a(t)}}
\ee

The three main predictions of the Big Bang theory are the following:
\begin{itemize}
\item
The expansion of the universe (Hubble's law).
\item
The existence of the Cosmic Microwave Background (CMB).
\item
The relative abundances of the light elements (nucleosynthesis).
\end{itemize}

A simple way to derive Hubble's law in the case of a flat (k=0) universe is to
consider two points in space separated by coordinate (comoving) distance $dr$
(with $d\theta=d\phi = 0$). The proper (physical) distance between the two
points
is given by
\be
ds= a(t) dr
\ee
Therefore, the physical relative velocity between the two points is
\be
{\dot s} = {\dot a} r + a {\dot r}
\ee
where ${\dot a} r$ is known as {\it the Hubble flow} and $a{\dot r}$ is the
{\it peculiar
velocity}. For large comoving distances $r$ the second term can be ignored  and
 we have
\be
v={\dot s} \simeq {{\dot a}\over a} s \equiv H(t) s \Rightarrow c z\simeq v =
H(t) s
\ee
where $z$ is the Doppler redshift and $H(t)\equiv {{\dot a}\over a}$ is the
Hubble
constant at cosmic time t. Eq. (13) is known as the {\it Hubble's
law}\cite{h27} and has been
verified by observations. The value of $H(t_0)$ may be written as
\be
H(t_0) = 100 h \hskip 2mm km/(sec \cdot Mpc)
\ee
with $h\in [1/2,1]$.

A particularly important prediction of the Big Bang theory is the existence of
the
CMB\cite{p88}. Penzias and Wilson\cite{pw65} observed for the first time in
1965 a highly isotropic
backgroung of microwave photons with spectrum that was thermal to a high
accuracy. Several
experiments since then have verified this observation and today it is known
that the
temperature of this background is given by Eq. (10) ($t=t_0$) and temperature
anisotropies are of
$O(10^{-5})$.
An additional term of {\it dipole} anisotropy \cite{fcw83} is known to be
present due to our motion
with respect to the CMB. The magnitude of this term is  ${{\delta T}\over T}
\simeq 10^{-3}$ and
corresponds to a velocity of the earth of about $600km/sec$ with respect to the
CMB frame. Even
though the dipole term dominates over the primordial fluctuations it can be
easily subtracted due
to its dipole nature.

 CMB photons are free at the present time $t_0$ (they have mean free path
of cosmological scale) and there is no apparent mechanism that could have
caused their
thermalization. However, such mechanism is naturally provided by the Big Bang
theory.
According to this theory, the universe is hotter at early times (cf Eq. (10))
and therefore
there is a time $t_{rec}<t_0$ ({\it time of recombination}) when photons are
energetic
enough to ionize matter $(T(t_{rec}) \gsim T_{ion} \simeq O(10^4) \hspace{1mm}
^0 K)$. In such a
primordial plasma, thermalization of photons can occur
effectively\cite{ah50,dprw65}.

The redshift $z(t_{rec})$ at the time of recombination defined as
\be
1+z(t_{rec}) \equiv {{a(t_0)}\over {a(t_{rec})}} = {{T(t_{rec})}\over {T(t_0)}}
\ee
is therefore $z(t_{rec})\simeq 1500$. Thus, according to the Big Bang theory
the CMB photons were scattered for last time at $t=t_{rec}$ and carry a
`photograph'
of the universe taken when it was 1500 times smaller and hotter. There is
exciting
information hidden in this map. An as yet unknown source has created
primordial fluctuations that evolved gravitationally into what we see as
galaxies,
clusters and large scale structure today. It is those fluctuations in their
primordial
gravitationally unaffected form that have been recorded by the CMB photons.
Clearly,
information on such a primordial pattern can impose severe constraints on
theories
attempting to explain the origin of primordial fluctuations.

The third prediction of the Big Bang theory is made on the relative
cosmological abundances of the light elements $(H,He,H^3, Li^7)$ which span
$10$
orders of magnitude and are predicted correctly by the Big Bang theory provided
that
the baryon to photon ratio is $\eta\equiv {{n_B}\over {n_\gamma}}\simeq
10^{-10}$\cite{abg48}.

The assumptions of homogeneity and isotropy made by the Big Bang theory, even
though
consistent with very large scale ($\gsim 100h^{-1} Mpc$)
observations\cite{ssotkl92} are only first
approximations to the realistic system. It is enough to look at the night sky
to
realize that the universe is not homogeneous and isotropic on relatively small
scales. Instead, matter tends to cluster gravitationally and form well defined
structures. In fact, detailed large scale structure observations have indicated
the
existence of {\it sheets} and {\it filaments} of galaxies separated by large
voids.
The typical scale of these structures is about $50 h^{-1} Mpc$
\cite{zes82,bs83,lgh86,dh82}.

Fig. 1 shows a slice of the CfA survey
which is a map of galaxies in redshift space (depth in the sky). The
presence of the above described structures is evident in this map (the
axis of redshifts can easily be converted to distance (depth in the
sky) by using Hubble's law\footnote{For example for $h=1$ dividing $z$
by $100$ gives the corresponding distance in $Mpc$.}). Can these large
structures be explained by assuming that matter moved due to gravity
from an initially homogeneous state?

{\bf Figure 1:} {\it The CfA Survey} shows a map of galaxies in redshift space
(depth in the sky).
This `slice of the universe' includes a range of about $6^0$ in declination
space.
\vspace{8cm}

The maximum distance that matter can have travelled since the Big Bang is
\be
\delta r_{max} \simeq \delta v \hspace{1mm} t_0 \simeq 6 h^{-1} Mpc
\ee
where $\delta v \simeq 600 km/sec$ is the typical peculiar velocity of galaxies
(induced by gravity) at the present time $t_0$. This $\delta r_{max}$ is much
less
than the typical scale of observed structures. Thus, these structures can not
be the
result of gravity alone. They must also reflect the presence of primordial
perturbations. The question that we want to address is {\it What caused these
fluctuations?}

Until about 15 years ago there was no physically motivated mechanism to cause
these
perturbations. However, during the past decade, two classes of theories
motivated
from quantum field theory have emerged and attempt to give physically motivated
answers to the question of the origin of structure in the universe.

A model for large scale structure formation is characterized by two basic
features.
The first is the kind  of the assumed dark matter and the second is the
type of primordial fluctuations.

Dynamical measurements based mainly on peculiar
velocities have shown that at least $90$\% of the matter in the universe is not
luminus\cite{t87,bd89,sdyh92}. This non-luminus matter which has not been
directly detected yet, is
called {\it dark matter} and its properties determine critically the rate of
gravitational
growth of primordial perturbations at different cosmic epochs. In particular,
dark
matter with low non-relativistic velocities at the time $t_{eq}$ (when
fluctuations
start to grow) is called cold dark matter (CDM)\cite{pss88}. A CDM candidate
motivated from
particle physics is the hypothetical particle {\it axion}\cite{pww83} which
appears to be a
necessary consequence of a successful resolution of the strong CP problem in
QCD.
Dark matter which has relativistic velocities at $t_{eq}$ is called hot dark
matter
(HDM) and a primary candidate for it is a massive $(25 h^{-2}eV)$ neutrino. Due
to
their relativistic velocities HDM particles can not cluster gravitationally at
early
times and small scales. This effect is called {\it free streaming}\cite{bes80}.
The critical scale
for HDM perturbation growth is the free streaming scale $l_{fs}(t)\simeq v_\nu
(t) t$
\ie the distance travelled by a HDM particle in a Hubble time $t$.
Adiabatic perturbations (those produced by inflation) on scales $l<l_{fs}$ are
erased on small
(galactic) scales
due to the effects of free streaming.

The second characteristic of large scale structure formation models is the type
of
primordial fluctuations. There are two broad classes of primordial
perturbations,
both produced by physically motivated mechanisms. The first includes gaussian
adiabatic perturbations produced\cite{p80,cm80,l80,s81} during a period of
exponential growth of the
universe known as {\it inflation}\cite{g81}. These perturbations may be
represented as a
superposition of plane waves with random phases \ie
\be
\delta \equiv {{\delta \rho}\over \rho} (x)\sim \sum_k |\delta_k | e^{i
\theta_k}
e^{ikx}
\ee
where ${{\delta \rho}\over \rho} (x)$ represents the primordial density
fluctuation
pattern and $\theta_k$ are random uncorrelated phases. By the central limit
theorem,
since $\delta$ is a superposition of an infinite number of uncorrelated random
variables, its probability distribution is gaussian \ie
\be
P(\delta)\sim e^{-\delta^2 /\sigma^2}
\ee
Fig. 2a shows schematically the features of such a gaussian perturbation
pattern.
\newpage

{\bf Figure 2:} Gaussian fluctuations may be represented as a superposition
of plane waves with random phases (a) while topological defect fluctuations
are represented by a superposition of seed functions (b).
\vspace{8cm}

The second class includes primordial perturbations produced by a superposition
of
localized seeds created during a phase transition in the early universe. Such
seeds
are known as {\it topological defects}\cite{k76,k80,v85,t89,vs94} and their
formation is predicted by
many (but not all) Grand Unified Theories (GUTs)\cite{r85}. The pattern of
topological defect
perturbations may be represented as a superposition of localized functions
(Fig.
2b) \ie
\be
\delta=\sum_i {{\delta \rho}\over \rho} (x-x_i)
\ee
and therefore, in general the probability distribution of $\delta$ is not
gaussian\footnote{For a not very large number of superposed seeds.})

Both inflationary and topological defect perturbations can be combined with
either
HDM or CDM for the construction of structure formation theories. In models
based on
gaussian adiabatic fluctuations with HDM, galactic scale fluctuations are
erased by free
streaming and therefore these structures can only form at later times by
fragmentation of larger objects. This is in contrast with observations
indicating
that galaxies formed earlier than larger structures. These models face also
other
conflicts with observations (\eg they violate CMB fluctuation constraints) and
have
therefore been placed into disfavor. On the other hand, the model based on
inflationary gaussian fluctuations with CDM has been well studied and makes
concrete
predictions that are in reasonable agreement with most types of observations
especially on small and intermediate scales. This model is currently the {\it
`standard model'} for structure formation\cite{bfpr84}.

Models based on topological defect perturbations combined with either hot or
cold
dark matter have not been as well studied but they have been shown to have
several
interesting features that make them worth of further investigation. Most of the
remaining of these talks will focus on one of the most interesting types of
topological defects {\it cosmic strings}\cite{z80,v81}. Three basic aspects of
cosmic string
physics will be discussed: their {\it formation, evolution and gravitational
effects}.

\section{Formation of Topological Defects}
\subsection{Kibble Mechanism}

Even though no particle corresponding to a scalar field has been discovered so
far,
the role of scalar fields in particle physics models is central. Indeed, scalar
fields provide a natural and simple mechanism to induce spontaneous symmetry
breaking
in gauge theories thus achieving two important goals: First maintain gauge
invariance
and renormalizability of these theories and second give mass to gauge bosons
thus
making them consistent with inducing short range interactions like the strong
and
electroweak forces. Symmetry breaking can be achieved by the use of scalar
field
potentials of the form
\be
V(\Phi)={\lambda \over 4} (\Phi^\dagger \Phi - \eta^2)^2
\ee
where $\Phi = [\Phi_1, ... , \Phi_N ]$ is a multiplet of scalar fields.
\newpage

{\bf Figure 3:} {\it The Kibble Mechanism:}
The evolution of a scalar field in a potential with discrete
minima ((a) left) leads to the formation of a domain wall in physical space
after the relaxation of the field in its minima ((a) right). When the potential
minimum (vacuum manifold) has the topology of a circle ($S^1$) a cosmic
string forms in physical space (b) while a monopole forms when the vacuum
manifold is a sphere ($S^2$) (c).
\vspace{12cm}

Consider for example a one component scalar field whose dynamics is determined
by the
double well potential of Fig. 3a. At early times and high temperature T the
field
$\Phi$ will have enough energy to span the whole range of the
potential\footnote{The
temperature dependence of the effective potential is ignored here since its
calculation is highly non-trivial in out of equilibrium systems like the one
discussed here.}
and go over the barrier between the two minima. As the universe cools and
expands
however, the energy of the scalar field will drop and it will eventually be
confined
in one of the two minima $(+\eta, -\eta)$. The choice of the minimum is
arbitrary. In fact, there will be neighbouring uncorrelated or causally
disconnected
regions where the choice of the minimum (vacuum) will be opposite (Fig. 3a). By
continuity of the field $\Phi$ there will be a surface separating the two
domains
where $\Phi = 0$ and therefore $\Phi$ will be associated with the
high energy of the local potential maximum. This surface of trapped energy
density
is known as a {\it domain wall} and is the simplest type of topological defect.
Domain walls form in theories where the vacuum manifold $M$ (minimum of
potential)
has more than one disconnected components. This fact is expressed in homotopy
theory\cite{s51}
as $\pi_0 (M) \neq 1$.

The mechanism described above for the formation of domain walls is known as
the {\it Kibble mechanism}\cite{k76,k80} and applies in a similar way to the
case of
other topological defects. Consider for example the case where the dynamics of
a two
component scalar is determined by the potential of Fig. 3b. After the field
relaxes to its vacuum, there will be (by causality) regions of the universe
where the
field $\Phi$ will span the whole vacuum manifold as we travel around a circle
in
physical space.
Topologically stable vortices can also form in systems with multiple
scalars\cite{pexis}
while  metastable vortices can even form in systems where the vacuum manifold
is not $S^1$.
In the later cases the stability is not topological by dynamical\cite{ponth}.
 By continuity of $\Phi$ there will be a point inside this circle where
$\Phi=0$. Such a point (and its neighbourhood) will be associated with high
energy
density due to the local maximum of the potential at $\Phi=0$. By extending
this
argument to three dimensions, the point becomes a line of trapped energy
density
known as the {\it cosmic string}. In general cosmic strings form in field
theories
where there are closed loops in the vacuum manifold $M$ that can not be shrank
to a
point without leaving $M$. In homotopy theory terms, we require that the first
homotopy group of the vacuum should be non-trivial \ie $\pi_1 (M)
\neq 1$. It may be shown that this is equivalent to $\pi_0 (H) \neq 1$ where
$H$ is
the unbroken group in the symmetry breaking $G\rightarrow H$ (M=G/H). As will
be
discussed below, the number of times the field $\Phi$ winds around the vacuum
manifold as we span a circle in physical space around the string is a
topologically
conserved quantity called the {\it topological charge} (or {\it winding
number}) and
its conservation guarantees the stability of the string.

In theories where the vacuum manifold is a sphere, the defect that forms by the
Kibble mechanism is pointlike and is known as the {\it monopole} (Fig. 3c). In
general
monopoles form in theories where the vacuum manifold has unshrinkable spheres
\ie
$\pi_2 (M)\neq 0$ (equivalent to $\pi_1 (H)\neq 1$). Monopoles, like strings
have
associated topological charge whose conservation guarantees their stability. As
will
be discussed later, most types of domain walls and monopoles are inconsistent
with the
standard Big Bang theory as they lead to quick domination of the universe by
defect
energy and subsequent early collapse. On the other hand, in the case
of strings there is a natural mechanism to effectively convert string energy to
radiation and thus prevent the string network from dominating the universe.

Topological defects similar to the ones described above can form not only in a
cosmological setup
but also in several condensed matter systems \cite{m79} like liquid
crystals\cite{g74}, $He^3$
\cite{sv87} and superconductors\cite{a57}.

\subsection{Non-local defects: Textures}

All the
defects discussed so far have two important properties:
\begin{itemize}
\item
The field configuration can not be continously deformed to the trivial zero
energy
vacuum where the field would point to the same direction everywhere in space.
\item
There is a well defined region in space where most of the defect energy is
localized.
It will be shown that the size of this domain is proportional to the inverse of
the
symmetry breaking scale $\eta$.
\end{itemize}
 From the second property it follows that the above described defects are {\it
localized
defects}. There is a second class of defects that have the first but not the
second
of the above properties. {\it Textures} belong to this later
class\cite{d87,t89a}.
\newpage

{\bf Figure 4:} The field configuration of a non-local {\it texture} in one (a)
two (b) and three (c) space dimensions.
\vspace{10cm}

Consider a global (no gauge fields) theory where the vacuum manifold is $S^1$
(a
circle). Consider also the one dimensional (fixed boundary conditions) scalar
field
configuration shown in Fig. 4a. This configuration is topological (can not be
deformed to the trivial vacuum) but is non-local (there is no core of energy
density). It is known as the one dimensional texture and in general it is
non-static.
It is straightforward to generalize the one dimensional texture to two
dimensions and
three dimensions by considering theories with $M=S^2$ and $M=S^3$ (Fig. 4b and
4c). In fact it may be shown (Derrick's theorem\cite{d64,r87}) that a rescaling
of the radial
coordinate $r$ by a scale factor $\alpha > 1$ $(r\rightarrow \alpha r)$ leads
to a
rescaling of the energy of the three dimensional texture by the inverse factor
$\alpha$ $(E\rightarrow E/\alpha)$ which implies that collapse and subsequent
unwinding is an energetically favored process provided that the topological
charge
is larger than a critical
value\cite{p92,psb92,bcl91,s93}.
 Even though the physics and observational effects of textures are particularly
interesting\cite{ts90,gst91,stpr91,cost91}
subjects they are outside of the scope of this
review and will not be further discussed here
(for a good review of textures see Ref.
\cite{t91}).

\subsection{Classical Field Theory}

In what follows I will focus on the cosmological effects of cosmic strings. The
simplest model in which strings involving gauge fileds can form\cite{no73} is
the Abelian Higgs
model involving the breaking of a $U(1)$ gauge symmetry. The Lagrangian of the
Abelian Higgs model is of the form
\be
{\cal L}= {1\over 2} D_\mu \Phi D^\mu \Phi - V(\Phi)+ {1\over 4} F_{\mu \nu}
F^{\mu\nu}
\ee
where $F_{\mu \nu} = \partial_\mu A_\nu - \partial_\nu A_\mu$, $D_\mu \equiv
\partial_\mu - i e A_\mu$ and $V(\Phi)= {\lambda \over 4} (|\Phi|^2 -
\eta^2)^2$,
($\Phi = \Phi_1 + i \Phi_2$). The field configuration corresponding to a
string\cite{no73,r87} in
this theory is shown in Fig. 3b. The topological stability of the string
(vortex in
two space dimensions) is due to the conservation of the topological invariant
(topological charge or winding number)
\be
m=\int_0^{2\pi} {{d\alpha}\over {d \theta}} d\theta
\ee
where $\alpha$ is an angular variable determining the orientation of $\Phi$ in
the
vacuum manifold which is $S^1$ (circle) in the Abelian Higgs model and $\theta$
is
the azimouthal angle around the vortex in physical space. For the string shown
in
Fig. 3b we have $m=1$ and the field winds once in the positive direction as we
span a
circle around the string.

The field configuration of a vortex may be described by the following ansatz:
\ba
\Phi &=& f(r) e^{i m \theta}\\
{\vec A} &=& {{v(r)}\over r} {\hat e}_\theta
\ea
with $f(0)=v(0)=0$, $f(\infty)=\eta$ and $v(\infty)=m/e$. The forms of $f(r)$
and $v(r)$ can be obtained numerically\cite{no73}
from the field
equations with the above ansatz. The energy of the vortex
configuration (energy of string per unit length) is
\be
\mu = {E\over L} = \int d^2 x (f'^2 +{{v^2}\over {r^2}} + {{(ev-m)^2}\over
{r^2}}f^2
+{\lambda \over 4} (f^2 -\eta^2)^2 )
\ee
Clearly, $v\neq 0$ is required for finite energy \ie the gauge field is needed
in
order to screen the logarithmically divergent energy coming from the angular
gradient of the scalar field. A vortex with no gauge fields is known as {\it a
global
vortex} and its energy per unit length diverges logarithmically with distance
from
the string core. This divergence however is not necessarily a problem in
systems where
there is a built-in scale cutoff like systems involving interacting vortices.
In such
systems the intervortex distance provides a natural cutoff scale for the energy
integral. Similar considerations apply to monopoles with no gauge fields even
though the energy
divergence in that case is linear rather than logarithmic\cite{bv89,rb90,plarg}

Even though no analytic solution has been found for the functions $f(r)$ and
$v(r)$
it is straightforward to obtain the asymptotic form of these functions from the
field
equations (Nielsen-Olesen equations).
The obtained asymptotic form for $r\rightarrow \infty$ is\cite{no73,r87} (but
see Ref. \cite{pasym}
for a correction to the standard result)
\ba
f(r) &\rightarrow & \eta - {{c_f}\over \sqrt{r}} e^{-\sqrt{\lambda} \eta r}\\
v(r) &\rightarrow &  {m\over e} -c_v \sqrt{r} e^{-e \eta r}
\ea
where $c_v$ and $c_f$ are constants.
Therefore the width of the vortex is
\be
w\sim \eta^{-1}
\ee
where $\eta$ is the parameter of the symmetry breaking potential also known as
{\it
the scale of symmetry breaking}.
The energy per unit length $\mu$ may also be approximated in terms of $\eta$ as
\be
\mu \simeq \int_w d^2 x V(0) \sim \eta^2
\ee
The typical symmetry breaking scale for GUTs is $\eta \simeq 10^{16} GeV$ which
leads to extremely thin and massive strings
\ba
w &\simeq & 10^{-30} cm\\
\mu &\simeq & 10^{14} tons/mm
\ea

Therefore, the Nielsen-Olesen vortex may be viewed as a linear, topologically
stable
field configuration involving both magnetic field energy and Higgs potential
energy that decay exponentially at large distances. No finite gauge
transformation
can gauge away a vortex gauge field everywhere in space. The several
interesting
quantum field theoretical interactions of vortices with particles (fermions)
are out of the
scope of this review\cite{pscat}.

\section{Cosmic String Evolution}
\subsection{Nambu Action}

 The exact analytic treatment of the evolution of a network of strings would
involve
the analytic solution of the time dependent nonlinear field equations derived
from the
Lagrangian of Eq. (21)  with
arbitrary initial conditions. Unfortunately no analytic solution is known for
these
equations even for the simplest nontrivial ansatz of the static Nielsen-Olesen
vortex.
Even the numerical solution of these equations is practically impossible for
systems of
cosmological scales and with complicated initial conditions.
The obvious alternative is to resort to realistic approximations in the
numerical
solution of the field equations in cosmological systems. First, the correct
initial
conditions for a numerical simulation must be obtained by simulating the above
described Kibble mechanism on a lattice. This may be achieved by implementing a
Monte-Carlo simulation implemented first by Vachaspati and Vilenkin\cite{vv84}.
They considered a discretization of the vacuum manifold ($S^1$) into
three points (Fig. 5) and then assigned randomly these three discrete
phases to points on a square lattice in physical space in three
dimensions. For those square plaquettes for which a complete winding of
the phase in the vacuum manifold occured, they assigned a string segment
passing through. It may be shown that this algorithm leads to strings
with no ends within the lattice volume \ie strings either form loops or
go through the entire lattice volume (infinite strings).
This simulation showed that the initial string network consists of 80\% long
strings
and 20\% loops.
\newpage

{\bf Figure 5:} The Vachaspati-Vilenkin algorithm used to study the formation
of a string network by Monte-Carlo simulations.
\vspace{8cm}

In order to find the cosmological effects of this initial string network, it
must be
evolved in time. Since it is impractical to evolve the full field equations on
the
vast range of scales $10^{-30} cm$ (string width) to $10^3 Mpc$ (largest
cosmological
scales) we must resort to some approximation scheme. The ratio $w/R$ of the
width of
the string over the string coherence scale is an intrinsically small parameter
for
cosmological strings ($w/R << 10^{-30}$ for $t>t_{eq}$). Thus, this parameter
can be
used to develop a perturbation
expansion\cite{f74,t88} for the action that
describes the dynamics of a string segment. The
zeroth order approximation is obtained for $w/R
= 0$.

A heuristic derivation of the action of a zero width string may be obtained by
using
the analogy with a relativistic free point particle with mass $m$ and velocity
$v$.
The relativistic action describing the motion of such a particle is of the form
\be
S=-m\int_A^B d\tau = -m \int_A^B dt (1-v^2)^{1/2} \simeq \int_A^B dt ({1\over
2} m v^2)
+ const
\ee
where $A$, $B$ are the endpoints of the trajectory and $\tau$ is the particle's
proper time. As shown in Eq. (32), the non-relativistic limit is obtained for
low
velocities.

The trajectory of the string is not parametrized by only one variable (the
proper
time $\tau$) but by two variables (a timelike variable $\tau$ and a spacelike
variable $l$ which may be viewed as parametrizing the length of the string).
Thus,
the generalization of the point particle relativistic action to the string with
mass
per unit length $\mu$ is the {\it Nambu action} defined as:
\be
S=-\mu \int_A^B dl d\tau \sqrt{-g^{(2)}} + O(w/R)
\ee
where $g^{(2)}= det(X_{,a}^\mu X_{,b}^\nu g_{\mu \nu})$  ($a,b = 1,2\rightarrow
(\tau,l)$, $\mu,\nu=1,...4$) is the metric of the world sheet spanned by the
string.
The Nambu action is an excellent approximation to the dynamics of
non-intersecting
cosmic string segments and is much simpler to handle numerically than the full
field
theoretic action.

\vspace{8cm}
{\bf Figure 6:} The intercommutation (exchange of string segments) occurs when
string segments intersect and favors formation of string loops (a). The effect
is closely related to the right angle scattering of vortices in head on
collisions (b,c) (from Ref. \cite{pinst}).

A crucial assumption in the derivation of the Nambu action is that string
segments do
not interact with each other. This assumption breaks down when two string
segments
intersect. The outcome of such an event can only be found by evolving the full
field
equations. Such numerical experiments have shown\cite{s87,m88,pinst} that at
intersections string
segments {\it exchange parteners} and a process known as {\it intercommuting}
occurs (Fig.
6a). Intercommuting is closely related to the right angle scattering of
vortices in
head on collisions (Fig. 6b,c) which has been observed in several numerical
experiments
and can be analytically understood using dynamics in moduli spaces\cite{r88}.
As shown in Fig.
6a intercommuting tends to favor the formation of string loops which oscillate
and decay
by emitting gravitational radiation with characteristic frequency $\omega \sim
R^{-1}$
($R$ is the loop radius) and rate ${\dot E} = \gamma G\mu^2$ ($\gamma =
const\simeq
50$)\cite{vv85}.

Therefore, this mechanism for loop formation, provides also an efficient way
for
converting long string energy that redshifts with the universe expansion as
$E_{str}
\sim a^{-2}$ to radiation energy that redshifts as $E_{rad} \sim a^{-4}$. This
is the
crucial feature that prohibits strings from dominating the energy density of
the
universe during the radiation era and makes the cosmic string theory a viable
theory
for structure formation in the universe.

\subsection{Scaling Solution}

Using the Monte-Carlo initial conditions described above, the Nambu action to
evolve
string segments and intercommuting to describe intersection events, it is
straightforward (though not easy) to construct numerical simulations describing
the
evolution of the string network in an expanding universe. Using this approach
it has
been shown\cite{at85} that the initial string network quickly relaxes in a
robust way to a scale
invariant configuration known as {\it the scaling solution}. According to the
scaling
solution the only scale that characterizes the string network is the horizon
scale at
any given time. On scales larger than the causal horizon scale $t$, the network
consists of a
random walk of long strings which are coherent on approximately horizon scales.
On
scales smaller than the horizon more recent simulations\cite{bb88,as90,at89}
(Fig. 7) have shown that
the network consists of a fixed number of approximately 10 long strings
coherent on horizon scales
and a large number of tiny loops with typical radius $10^{-4} t$. The efficient
formation of tiny
loops by the intercommuting of long strings on small scales leads to the
existence of wiggles on
the long strings (wiggly strings). The {\it effective} mass per unit length of
wiggly strings is larger than the bare mass ($\mu_{eff} \simeq 1.4 \mu$) and
their
tension is smaller than the bare tension. The main features of wiggly
strings\cite{c90,cka92,v92,vv91} are discussed in more detail below.
\newpage

{\bf Figure 7:}  The string network in the matter era.
The volume of the cube is $(H/2)^3$ and the simulation box was plotted after an
expansion by a
factor $a=16$ (from Ref. \cite{bb88}).
\vspace{8cm}

There is a simple heuristic way to understand why does the network of strings
approach a scaling solution with a fixed number of long strings per horizon
scale.
Consider for example the case when the number of long strings per horizon
increases
drastically. This will inevitably lead to more efficient intercommuting and
loop
formation thus transfering energy from long strings to loops and reducing the
number
of long strings per horizon back to its equilibrium value. Similarly if the
number of
long strings decreases reduced intercommuting and loop formation will tend to
increase
the number of long strings towards an equilibrium value. These heuristic
arguments
have been put in more detailed form using differential equations in Ref.
\cite{v85}.

\subsection{Benefits of Scaling}

The scaling behavior described above is a crucial feature of cosmic strings. It
ensures that the string energy density remains a fixed fraction of the matter
density
and thus an overclosure and premature collapse of the universe is avoided. To
see this
we may calculate the total energy in long strings at a time $t$ assuming a
fixed number
of $M$ long strings per horizon volume. The energy density of strings is
\be
\rho_{str} \sim {{M \mu t}\over t^3} \sim {\mu \over t^2} \sim G\mu \rho_c
\ee
where I used the definition of the critical density $\rho_c \sim ({{\dot a}
\over
a})^2$ (see Eq. (3) with $k=0$). Eq. (34) implies that
\be
{{\rho_{str}}\over {\rho_c}}\simeq G\mu \simeq 10^{-6}
\ee
for $\eta = 10^{16} GeV$. Therefore the density in strings remains a small
fixed fraction of the
energy density of the universe. As will be seen, this nice feature is not
shared by domain
walls even if it is assumed that walls approach a scaling solution.

Gauge monopoles and domain walls are inconsistent with standard cosmology each
for a
different reason. In the case of gauge monopoles there is no long range
interaction
and therefore there is no efficient mechanism to convert energy from monopoles
into
radiation. The monopole-antimonopole anihilation for example is very
inefficient
without long range interactions. Therefore, during the radiation era monopole
energy
redshifts with radiation as $\rho_{mon} \sim a^{-3}$ and ${{\rho_{mon}} \over
{\rho_{rad}}}$ scales as $a(t)$ leading to $\rho_{mon}(t_0) >> \rho_c (t_0)$
and a
premature collapse of the universe\cite{zk78}.

If Grand Unification to a simple group is realized
in nature, the formation of monopoles is inevitable. This is the well known
monopole
problem of standard cosmology and may be seen as follows: Consider the symmetry
breaking $G\rightarrow H$ of a simple GUT simple group
$G$ to a group $H$. In any realistic theory, $H$ must include the gauge group
of
electromagnetism $U(1)_{em}$ since it is an experimental fact that the photon
is
massless and therefore $U(1)_{em}$ is unbroken.
 Now, the vacuum manifold at the broken symmetry phase is $M=G/H$. The
homotopy sequence of homotopy theory may be used to show that
\be
\pi_2 (G/H) = \pi_1 (H) = \pi_1 (U(1)_{em}) = Z
\ee
and therefore since $\pi_2 (M) \neq 1$ monopoles must form in GUTs.

The dilution of the monopole density during an epoch of exponential expansion
of the
universe (inflation) provides one solution to the monopole problem\cite{g81}.
An alternative\cite{lp80}
solution is provided by constructing GUT models where monopoles get temporarily
connected by $U(1)_{em}$ strings during a temporary breaking of
electromagnetism.
Such models have the disadvantage of being somewhat unatural by introducing
extra
phase transitions in the symmetry breaking sequence of GUTs.

GUT domain walls are inconsistent with standard
cosmology\cite{zk74} even if they manage to
achieve a scaling solution. Consider for example
a domain wall network with a single wall
spanning each horizon scale. The wall energy
density may be easily found as
\be
\rho_w \sim {{V(0) \eta^{-1} t^2}\over {t^3}} \sim \eta^3 t^{-1} \sim
{{(G\eta^2)
\eta t}\over {G t^2}}
\ee
Thus for $t=t_0$ and $\eta\simeq 10^{16} GeV$ we obtain
\be
{{\rho_w}\over {\rho_c}} (t_0) \simeq 10^{52}
\ee
The wall symmetry breaking scale must be at least $17$ orders of magnitude
smaller than the GUT scale for a model containing walls to be viable without
inflation.

The above discussion shows the unique attractive features of cosmic strings
compared
to other defects in the context of standard cosmology. Several other such
features
will be discussed in the rest of this review.

\section{Gravitational Effects}
\subsection{String Metric: The Deficit Angle}

The most important interaction on cosmological scales is gravity. It is
therefore
important to understand the gravitational effects of strings before attempting
to
study in more detail their cosmological effects.
The straight string solution is thin, cylindrically symmetric and Lorentz
invariant for
boosts along the length of the string. This imposes the following constraint on
components of the energy momentum tensor $T_{\mu\nu}$
\be
T_z^z (\rho) = T_0^0 (\rho) \simeq \mu \delta (x) \delta (y)
\ee
Also
\be
T^\nu_{\mu,\nu} =0 \Rightarrow {d\over {dx}} T_x^x (\rho) = 0 \Rightarrow
T_x^x=T_y^y=0
\ee
where use was made of the cylindrical symmetry and of the fact that $T_x^x =
T_y^y
\rightarrow 0$ as $r\rightarrow \infty$. Therefore, the string energy momentum
tensor
may be approximated by
\be
T_{\mu \nu} \simeq \delta (x) \delta (y) diag(\mu,0,0,-\mu)\equiv
diag(\rho,p_x,p_y,p_z)
\ee
which implies that the string has significant negative pressure (tension) along
the z
direction \ie
\be
p_z = -\rho =-\mu \hspace{1cm} p_x=p_y = 0
\ee
This form of $T_{\mu\nu}$ may be used to obtain the Newtonian limit for
gravitational
interactions of strings with matter. For the Newtonian potential $\Psi$ we have
\be
\nabla^2 \Psi = 4\pi G(\rho + \sum_i p_i)=0 \Rightarrow {\vec F_N} = 0
\ee
and a test particle would feel no force by a nearby motionless straight string.

Simulations have shown however that realistic strings are neither straight nor
motionless. Instead they have small scale wiggles and move with typical
velocities of
$v_s\simeq 0.15c$ coherent on horizon scales. What is the energy momentum
tensor and
metric of such wiggly strings?

The main effect of wiggles on strings is to destroy Lorentz invariance along
the
string axis, to reduce the effective tension and to increase proportionaly the
effective mass per unit length of the string. Thus, the energy momentum tensor
of a
wiggly string is
\be
T_{\mu \nu} \simeq \delta (x) \delta (y) diag(\mu_{eff}, 0, 0, -T)
\ee
with $T\equiv -p_z <\mu$, $\mu_{eff} > \mu$ and $\mu_{eff} T = \mu^2$
\cite{vv91} ($\mu$ is the
`bare' mass per unit length obtained from the field Lagrangian). As expected
the
above $T_{\mu \nu}$ reduces to the straight string case for $\mu_{eff} = T$.
The breaking of Lorentz invariance by the wiggles also induces a non-zero
Newtonian
force between the wiggly string at rest and a test particle since the tension
(negative pressure) is not able to completely cancel the effects of the energy
density
(Eq. (43)). This may be seen more clearly by using the Einstein's equations
with
the tensor of Eq. (44) to find the metric around a wiggly string. The result in
the weak
field limit (small $G\mu$) is\cite{vv91}
\be
ds^2 = (1+h_{00}) (dt^2 -dz^2 - (1-4G\mu_{eff})^2 r^2 d\varphi^2)
\ee
with
\be
h_{00} = 4G(\mu_{eff} - T) ln(r/r_0)
\ee
where $r_0$ is an integration constant. Clearly, in the presence of wiggles
($\mu_{eff} \neq T$) the Newtonian potential $h_{00}$ is non-zero.
The change of the azimouthal variable $\varphi$ to the new variable $\varphi'
\equiv
(1-4G\mu_{eff})\varphi $ makes the metric (45) very similar to the Minkowski
metric
with the crucial difference of the presence of the Newtonian $h_{00}$ term and
the
fact that the new azimouthal variable $\varphi'$ does not vary between $0$ and
$2\pi$ but between $0$ and $2\pi - 8\pi G\mu_{eff}$. Therefore there is a {\it
deficit
angle}\cite{v81a} $\alpha = 8\pi G \mu_{eff}$ in the space around a wiggly or
non-wiggly string.
Such a spacetime is called {\it conical} (Fig. 8) and leads to several
interesting
cosmological effects especially for moving long strings.

\vspace{10cm}
{\bf Figure 8:} {\it The gravitational effects of a long string:}
The string deficit angle $\alpha$ leads to sharp discontinuities in the
temperature of the CMB,
velocity fluctuations in matter and formation of wakes, and lensing of galaxies
and quasars.
The Doppler and Sachs-Wolfe effects (see section 6.2) due to plasma velocities
and
potential fluctuations on the last scattering surface ($t_{rec}$) are also
illustrated.

\subsection{Velocity Fluctuations}

The main mechanism by which strings create perturbations that could lead to
large
scale structure formation is based on velocity perturbations created by moving
long
strings.
 Long, approximatelly straight strings moving with velocity
$v_s$ induce velocity perturbations to the surrounding matter directed towards
the
surface swept in space by the string. This effect may be seen more clearly by
using
cartesian coordinates\cite{vv91}.

Consider a straight long straight long string moving on the $y-z$ plane with
velocity
$v_s$ (Fig. 8). By transforming the line element of Eq. (45) to `shifted'
cartesian
coordinates  defined as
\ba
y' &\equiv & r \sin \varphi' \simeq y - r \cos \varphi \hskip 1mm 4G\mu_{eff}
\varphi
\\
x' &\equiv & r \cos \varphi' \simeq r \cos \varphi = x
\ea
we obtain
\be
ds^2 = (1+h_{00}) (dt^2 -dz^2 -dx^2 -dy'^2)
\ee
where $y' \equiv y -4G\mu_{eff} \varphi x$. The geodesic equations for a test
particle in the cosmic string spacetime are of the form
\ba
2 {\ddot x} &=& - (1- {\dot x}^2 - {\dot y}'^2) \partial_x h_{00} \\
2 {\ddot y'} &=& - (1- {\dot x}^2 - {\dot y}'^2) \partial_y h_{00}
\ea
Using the form of $h_{00}$ from Eq. (46) and perturbing around the initial
particle
trajectory (on the string frame)
\be
x=v_s t \hspace{1cm} y'=y_0 = const
\ee
the geodesic equations become
\ba
{\ddot x} &=& -{1\over 2} (1-v_s^2) 4 G (\mu_{eff} -T) {x\over {r^2}} \\
{\ddot y'} &=& -{1\over 2} (1-v_s^2) 4 G (\mu_{eff} -T) {{y'}\over {r^2}}
\ea
A heuristic solution to these equations may be obtained as follows\cite{vv91}:
The Newtonian
force induced by the wiggly string on a particle at distance $r$ has a
magnitude (cf Eq. (53-54))
\be
F={{2mG (\mu_{eff} -T)}\over {\gamma_s r}}
\ee
and is effective for approximatelly
\be
\Delta t \simeq {r\over {v_s}}
\ee
Therefore the magnitude of the induced velocity will be approximatelly
\be
{\dot y}' \simeq {F\over m} \Delta t \simeq {{2G (\mu_{eff} -T)}\over {\gamma_s
v_s}}
\ee
and by symmetry it is directed towards the surface swept by the string.
The exact solution to the system of Eq. (53-54) turns out to be very similar
\be
{\dot y}' \equiv {\dot y} - 4\pi G\mu_{eff}v_s \gamma_s
\simeq {{2\pi G (\mu_{eff} -T)}\over {\gamma_s v_s}}
\ee
and therefore the total velocity perturbation induced by a moving long string
to
surrounding matter close to the surface swept by the string  is
\be
\Delta v = {{2\pi G (\mu_{eff} -T)}\over {\gamma_s v_s}} + 4\pi G \mu_{eff} v_s
\gamma_s
\ee
where the first term is due to the Newtonian interaction of the string wiggles
with
matter while the second term is an outcome of the conical nature of the
spacetime.

\subsection{Microwave Background Fluctuations}

Another particularly interesting effect of the string induced deficit angle is
the
creation of a characteristic signature on CMB photons.
The type of this signature may be seen in a heuristic way as follows:
 Consider a straight long
string moving with velocity $v_s$ between the surface of last scattering
occuring at
$t_{rec}$ and an observer at the present time $t_0$ (Fig. 8). The metric around
the string is approximated by Eq. (49) and at large distances from the string
core we have
\be
dy' \simeq dy -4G\mu_{eff} \varphi dx = dy - 4G\mu_{eff} \varphi v_s \gamma_s
dt
\equiv dy - V(\varphi) dt
\ee
 For photons we have $ds^2 =0$ and therefore the Newtonian
term $h_{00}$ has no effect. Thus the presense of
the moving long string between the observer and the last scattering surface
induces an
effective Doppler shift to the CMB photons. For photons reaching the observer
through
the `back' (`front') of the string we have
\be
({{\delta T}\over T})_{1,(2)} = ({{\delta v}\over v})_{1,(2)} = \pm(V(\varphi =
2\pi) -V(\varphi
=\pi)) = \pm 4\pi G \mu_{eff} v_s \gamma_s
\ee
where the $1 (2)$ and $+ (-)$ refer to photons passing through the `back'
(`front')
of the string.

Therefore a moving long string present between $t_{rec}$ and today induces line
step-like discontinuities on the CMB sky with magnitude\cite{ks84,g85}
\be
{{T_1 - T_2} \over T} = 8\pi G \mu_{eff} v_s \gamma_s
\ee
Notice that there is no Newtonian term inversely proportional to the wiggly
string
velocity as was the case for the induced velocity perturbations. In section 6
it
will be shown that Eq. (61) can be used to construct a superposition of long
string
perturbations thus constructing approximations to the predicted CMB maps.

\subsection{Lensing}

The existence of the deficit angle in the cosmic string space-time implies
that strings act as gravitational lenses with certain characteristic
properties\cite{v81a,g85,hn84}. A long string present between a quasar and an
observer
(Fig. 8) will lead to the formation of double quasar images
for the observer. The separation angle of the two images depends on
the magnitude of the deficit angle and may be found by simple
geometrical considerations. Using Fig. 8 the separation angle $\Delta \theta$
is obtained as
\be
(l+d) \sin ({{\Delta \theta}\over 2}) =  \sin ({\alpha \over 2})
\Rightarrow \Delta \theta \simeq 8 \pi G \mu_{eff} {{l}\over {l+d}} \lsim
5"
\ee
 Therefore a gravitational lensing event induced by a cosmic string is
expected to involve a number of neighbouring double images with typical
separation of a few arcsec. Such a candidate event has indeed been observed
and it will be discussed in some detail in section 5.5.

\section{Structure Formation}
\subsection{Sheets of Galaxies, Filaments}

 In the cosmic string model for structure formation, small and intermediate
scale structure (galaxies and clusters) is seeded by string loops while
large scale structure is produced by perturbations induced by long strings.

A loop with radius $R$ has a mass $M_l= \beta R \mu$ where $\beta$ is a
parameter approximatelly equal to $2\pi$ and on distances much larger
than its radius it produces a gravitational field which is identical
to the field of a point mass with mass $M_l$\cite{t84}. Therefore string loops
can act as seedlike perturbations leading to the formation of
clusters and galaxies. Recent simulations\cite{bb88,as90} however have shown
that
the typical size of loops is probably too small to have a significant
effect on the formation of objects like galaxies or clusters. This
implies that the simple
`one loop one object' assumption that was made in the
early days of the cosmic string model is probably incorrect
and more sophisticated methods must be developed to study the
way small and intermediate structure forms in the string model.

 The indicated relatively small importance of loops further amplifies
the cosmological role of long strings. These strings can lead to large
scale structure formation through the velocity perturbations, produced
during their motion, to surrounding matter\cite{v86}. As discussed in section 4
long strings moving with velocity $v_s$ produce velocity perturbations
directed towards the surface they sweep in space. These perturbations
whose initial magnitude is given by Eq. (59), grow gravitationally and
within approximatelly a Hubble time $t$ they form planar density perturbations
called wakes\cite{sv84,v86,svbst87,bps90,hm93} (Fig.
8). The gravitational growth of wakes can be calculated analytically in the
linear regime using a
simple but powerful method known as the {\it the Zeldovich approximation}.
 Using this method the thickness and typical dimensions of the
dominant predicted planar structures can be calculated\cite{bps90} and the
result can
be compared with observations of redshift surveys in order to test
the cosmic string model. In what follows I will sketch the basic steps of
the calculation involving the Zeldovich approximation.

Consider a long straight string moving with velocity $v_s$ and sweeping a
plane in an expanding universe. Consider also a test particle located
a physical distance $h(t)$ from the plane swept by the string. The
scale $h(t)$ will initially grow with the universe expansion but due
to the string induced velocity perturbation the growth will be
decelerated by gravity, the scale $h(t)$ will stop expanding,
it will turn around and collapse on the string induced wake.
 In order for the planar structure formed to be able to fragment and
lead to the formation of galaxies and clusters within it, it is
necessary that the developed overdensity be nonlinear \ie
${{\delta \rho} \over \rho}\gsim 0$. It may be shown that this
condition is indeed realized within the scales $h_{nl} (t)$ that
have turned around at a given time $t$. Thus, the scale $h_{nl} (t)$
that turns around at $t$ is called {\it the nonlinear scale} at time
$t$ and the time when a scale $h$ turns around to collapse is called
{\it the time of non-linearity} $t_{nl} (h)$ for the scale $h$. Clearly,
$h_{nl} (t)$ defines the thickness of the planar structure at the time
$t$. The length of this structure is approximated by the
coherence length of the string which is given by the horizon scale
$t$ while its width is approximated by the distance $v_s t$ travelled by
the coherent portion of the string within an expansion time $t$.

Observations indicate that the typical thickness of the observed sheets of
galaxies
is about $5h^{-1} Mpc$. By demanding $h_{nl} (t_0, G\mu) = 5h^{-1} Mpc$ \ie
that the
predicted thickness of string induced sheets of galaxies is equal to the
observed,
the single free parameter of the model $G\mu$ can be fixed. This normalization
can then
be compared with others coming from other observations and also from
microphysical
constraints. The length and width of the dominant structures can also be
obtained.

In the context of the Zeldovich approximation the physical (Eulerian)
coordinate
${\vec r}$ of a test particle with initial comoving position ${\vec q}$
(Lagrangian
coordinate) is written as
\be
{\vec r} ({\vec q},t) = a(t) ({\vec q} -{\vec \Psi} ({\vec q},t))
\ee
where ${\vec \Psi}$ is the comoving displacement induced by the initial
perturbation.
In order to find the comoving scale $q_{nl} (t)$ that turns around at time $t$
we
must first use dynamics to find ${\vec \Psi}$ and then solve ${\dot {\vec r}} =
0$ to
find $q_{nl} (t)$. This calculation is outlined below.

The dynamics of the Eulerian coordinate ${\vec r}$ may be obtained in the
Newtonian
approximation using the equations
\ba
{\ddot {\vec r}} ({\vec q},t) &=& - {\partial \over {\partial {\vec r}}} \Phi
({\vec r},
t)\\
{{\partial^2}\over {\partial {\vec r}^2}} \Phi ({\vec r},t) &=& 4\pi G \rho
({\vec
r},t)
\ea
In addition, mass conservation implies that the mass in a Eulerian volume $d^3
r$
should be equal to the mass in the corresponding Lagrangian volume $a^3 d^3 q$
\ie
\be
\rho ({\vec r}, t) d^3 r= a(t)^3 \rho_0 (t) d^3 q
\ee
or
\be
\rho ({\vec r}, t) = a(t)^3 \rho_0 (t) {1\over {det |{{d{\vec r}} \over {d{\vec
q}}}|}} \simeq \rho_0 (t) (1+ {{\partial }\over {\partial {\vec q}}} {\vec
\Psi} ({\vec
q},t))
\ee
{}From Eq. (66) and Eq. (68) we obtain to linear order in ${\vec \Psi}$:
\be
{\partial \over {\partial {\vec r}}} \Phi ({\vec r},t)\simeq 4\pi G [{1\over 3}
\rho_0
(t) {\vec r} + \rho_0 (t) a(t) {\vec \Psi} ({\vec q},t)]
\ee
Using the Friedman equation Eq. (3) in Eq. (69) and using also Eq. (64-65) we
obtain
\be
{\ddot {\vec \Psi}} + 2{{\dot a}\over a}{\dot {\vec \Psi}} + 3 {{\ddot a} \over
a}
{\vec \Psi} = 0
\ee
which with the appropriate initial conditions can determine the evolution of
${\vec
\Psi}$.

The initial conditions corresponding to the velocity perturbation induced by a
moving
long string at an initial time $t_i$ are of the form
\ba
{\vec \Psi} (t_i) &=& 0\\
{\dot {\vec \Psi}}_x (t_i)&=&{\dot {\vec \Psi}}_y (t_i)= 0\\
\Delta v & \equiv & a(t_i) {\dot {\vec \Psi}}_x (t_i)= 4\pi G \mu v_s \gamma_s
f
\ea
with $f=1+{{1-{T\over {\mu_{eff}}}}\over {2(v_s \gamma_s)^2}}$ as implied by
Eq. (59).

 The growing mode solution of Eq. (70) with initial conditions (71-73) in the
matter
era ($a\sim t^{2/3}$) can easily be found to be of the form
\be
\Psi (t) \simeq {3\over 5} \Delta v ({t\over {t_i}})^{2/3} ({{t_0}\over
{t_i}})^{2/3} t_i
\ee
 It is now straightforward to find the comoving scale $q_{nl}(t)$ that
becomes nonlinear and turns around at time $t$
\be
{d\over {dt}} (a (q_{nl} -\Psi (t)) =0 \Rightarrow q_{nl} (t) = 2 \Psi (t)
\ee
Using also the observational fact that $q_{nl} (t_0) \gsim 5h^{-1} Mpc$
we may obtain a constraint on the only free parameter of the model
\be
G\mu_{eff} \gsim 0.7 (v_s \gamma_s f)^{-1} \times 10^{-6}
\ee
where $t_i = t_{eq}$ has been used in Eq. (74) since fluctuations can
not grow before $t_{eq}$ due to pressure effects present in the
radiation component.
This result compares favourably with constraints coming from a completely
different direction: microphysics. In order for GUTs to be consistent with
low energy experiments and with constraints on baryon lifetime, the
scale of GUT symmetry breaking $\eta$ must be of the order $10^{16}$GeV.
For strings formed during a GUT phase transition this implies that
\be
(G\mu)_{GUT} \simeq (G\eta^2)_{GUT} \simeq 10^{-6}
\ee
which compares well with the corresponding constraint from macrophysics
(Eq. (76)). This striking agreement, which was first observed in studies of
galaxy formation by
string loops\cite{tb86}, between constraints coming from completely different
directions is an
exciting example of a meeting point between cosmology and particle physics and
has also been one of
the most attractive features of the cosmic string model.

 The earliest string wakes forming at $t_{eq}$ are the most numerous
and also they have the longest time to grow. Therefore they give rise to
the dominant sheetlike structures by today. Their typical dimensions
are determined by the size of the comoving horizon at $t_{eq}$ and by
the nonlinear scale $q_{nl} (t_{eq}, t_0)$ which determines their
present thickness (Eq. (75)).
 Thus the predicted dimensions of these structures are
\be
\xi t_{eq}^{com} \times v_s t_{eq}^{com} \times q_{nl} (t_{eq},t_0)
\simeq \xi 40 \times v_s 40 \times 5 Mpc^3
\ee
where $\xi \lsim 1$ is the coherence length of long strings.
The above predicted dimensions compare reasonably well with
observations assuming relativistic strings.  Simulations
have shown that a large portion of long strings have
relatively small velocities $v_s \simeq 0.15$
on horizon scales. Such strings will tend to form structures that are
more filamentary than sheetlike.

 The above discussion on the form and dimensions of the predicted
large scale structure has not taken into account the effects of
free streaming and therefore it is valid only for CDM. The effects
of free streaming present in the case of HDM can be taken into
account by preventing the growth of scales smaller than the
comoving free
streaming scale
\be
\lambda_j^{com}= v_\nu (t) t /a(t)
\ee
where the velocity $v_\nu$ of HDM particles (\eg neutrinos) starts
droping like $1/a(t)$ after $T\simeq m_\nu$ when they become
nonrelativistic.
 A sketch of the time dependence of $\lambda_j^{com}$ is shown in
Fig. 9.

\vspace{6cm}
{\bf Figure 9:} The time dependence of the free streaming scale
$\lambda^{com}_j$.

 An important value for $\lambda_j$ is its maximum value $\lambda_j^{max}$
which for adiabatic fluctuations produced during inflation
determines the minimum scale that can form independent of
larger scales (without fragmentation). Adiabatic (but not seedlike)
perturbations on all scales smaller than $\lambda_j^{max}$ are erased
by free streaming and structures on these scales can only form by
fragmentation of larger objects. This is a problem in adiabatic perturbations
with HDM because observations indicate\cite{wfd83} that galactic and cluster
scales formed earlier than larger scales. This is difficult to
explain in models where fragmentation is the only mechanism to form
smaller scale structures.
Seeds like cosmic strings
survive free streaming and therefore smaller scale fluctuations
in models with seeds + HDM are not erased but their growth
is only delayed by free streaming\cite{bkst87,bkt87}. Thus galaxies and
clusters
can in principle form independently of large scale structure in
these models.
 For a neutrino with mass $m_\nu =25eV$ (enough to produce $\Omega=1$
for $h=1/2$) we find
\be
\lambda_j^{max}  \simeq \lambda_j (t_{eq})\simeq 6 h_{50}^{-2} Mpc
\ee
and therefore the growth of all scales less than $6h_{50}^{-2} Mpc$
is delayed by free streaming.
 Thus the introduction of HDM in the string model has two main effects.
First it delays the growth of smaller scales thus transferring more power
on large scales and second the delayed growth tends to increase the
required value of $G\mu$ for nonlinear structures to form by today.
A detailed calculation shows\cite{bps90} that
\be
(G\mu)_{HDM} \simeq 2\times 10^{-6} > (G\mu)_{CDM}
\ee

\subsection{Peculiar Velocities}

Using the result of Eq. (74) for the comoving displacement $\Psi$ it is
straightforward to find
the magnitude and coherence length of peculiar velocities $u(t_0,t_i)$ produced
by long
strings at an initial time $t_i$ as observed at the present time $t_0$. The
typical
magnitude of these velocities is\cite{v92a}
\be
u(t_0,t_i)= {\dot \Psi} (t_0,t_i) = {2\over 5} \Delta v ({{t_0}\over
{t_i}})^{1/3}
\hspace{1cm} (t_i \geq t_{eq})
\ee
where $\Delta v \equiv 4 \pi G\mu_{eff} v_s \gamma_s f$. The coherence length
of
these velocity fields is given by the coherence length of the long string that
produced them which in turn is about equal to the comoving horizon scale
$L(t_i)\equiv t_i^{com}=\int_{t_i} {{dt}\over a(t)} \sim t_i^{1/3}$. Thus the
magnitude of the predicted velocity field with coherence scale $L$ is
\be
u(t_0,L)\simeq 300 \mu_6 h {{L_{eq}}\over L} km/sec
\ee
where $\mu_6 \equiv G\mu/10^{-6}$, $L_{eq}\equiv t_{eq}^{com}=14 h^{-2} Mpc$.
 This result is {\it not} in good agreement with peculiar velocity observations
on
large scales which indicate the presence of velocity fields with magnitude of
about
$600 km/sec$ coherent on all scales from $10h^{-1}Mpc$ to larger than $50
h^{-1}Mpc$.
 Even though it is possible to play with the normalization of the model to
induce
agreement with observations on a particular scale, the $L^{-1}$ scaling does
not
allow agreement on a large range of scales. This problem of the cosmic string
model
which also appears in the standard adiabatic CDM model can be addressed by
assuming
velocity bias which essentially means that the observed velocity fields do not
represent accurately the underlying velocity fields of dark matter since
luminous
matter evolves differently than dark matter. The motivation for such a
conjecture
however is not particularly appealing especially on these large scales.

\subsection{Power Spectrum}

An important quantity that characterizes a pattern of fluctuations ${{\delta
\rho}\over \rho} ({\vec x})$ is the {\it power spectrum}.  Consider a Fourier
expansion of a fluctuation pattern
\be
{{\delta \rho}\over \rho} ({\vec x})=\int d^3 k {{\delta \rho}\over \rho}
({\vec k})
e^{i {\vec k} {\vec x}}
\ee
The power spectrum $P({\vec k})$ of the pattern is then defined as the ensemble
average
\be
P({\vec k}) \equiv < |{{\delta \rho}\over \rho} ({\vec k})|^2 >
\ee
It may be shown that the correlation function $C({\vec x})$ of the pattern
smoothed
on a scale $l_0 \sim k_0^{-1}$ can be obtained as the Fourier transform of the
power
spectrum \ie
\be
C({\vec x}) \equiv <{{\delta \rho}\over \rho} ({\vec x_1}){{\delta \rho}\over
\rho}
({\vec x_1}+{\vec x})>_{x_1} = \int d^3 k P({\vec k}) e^{i {\vec k} \cdot {\vec
x}}
W(k-k_0)
\ee
where $W(k-k_0)$ is a {\it filter function} which filters out the scales that
do not
contribute due to smoothing and finite volume effects.
 Thus, the contribution of scales smaller than $l_0 \sim k_0^{-1}$ to the
variance of
the fluctuations is approximated by
\be
<({{\delta \rho}\over \rho} ({\vec x}))^2>_{l_0} = C(0)_{k_0}\equiv \delta
(k_0,t)^2
\simeq k_0^3 P(k_0)
\ee
Now the existence of the scaling solution implies that string induced
perturbations
$\delta_s$ have a fixed magnitude on horizon scales at any given time. Thus
\be
\delta_s (\lambda_k =t,t) =constant
\ee
where $\lambda_k = a(t) {{2\pi} \over k}$ is the physical smoothing scale.
At a later time $t$ in the matter era the fluctuations will have grown to
\be
\delta (k,t) = ({t\over {t_i}})^{2/3} \delta (\lambda_k =t_i,t_i)
\ee
But
\be
t_i=\lambda_k\sim a(t_i)/k \sim t_i^{2/3}/k\Rightarrow t_i\sim k^{-3}
\ee
and therefore combining Eq. (88) and Eq. (89) we obtain $\delta (k,t) \sim k^2$
which combined
with Eq. (87) gives
\be
P(k) \sim k
\ee
known as the {\it scale invariant\cite{h70} Harisson-Zeldovich power spectrum}.
This is a
generic type of spectrum which is also predicted by models based on
inflation\cite{gp82}. Any
large variation from the scale invariant spectrum is cosmologically
unacceptable since
it would lead to ultraviolet or infrared disasters in the magnitude of
fluctuations.

\vspace{8cm}
{\bf Figure 10:}   The $log$ of the power spectrum vs the $log$ of the scale
for strings (dashed)
and gaussian scale invariant fluctuations. In each case the curve with more
power on small scales
corresponds to CDM, the other to HDM. (From Ref. \cite{as92b}).

Given that growth of fluctuations starts at $t_{eq}$, perturbations on scales
larger
than the horizon at $t_{eq}$ enter the horizon after $t_{eq}$ and they keep
growing without delay thus keeping their power law form $P(k)\sim k$. On the
other
hand fluctuations on smaller scales enter the horizon at earlier times and
their
growth is delayed until $t_{eq}$ when matter dominates. The delay is larger the
earlier the scale enters the horizon \ie is larger for smaller scales. Thus, we
expect a {\it bending} of the spectrum at smaller scales\cite{as92a} to a power
low $P(k) \sim
k^n$ with $n<1$. An additional bending is expected to occur in the case of
HDM\cite{as92b} due to
the effects of free streaming which delay (erase) seed (adiabatic) fluctuations
on
small scales. In fact in the case of adiabatic fluctuations we have not only a
bending but a {\it cutoff} of the spectrum at the maximum free streaming scale.
The
above discussion is illustrated in Fig. 10 where the spectra of adiabatic and
string perturbations are plotted for CDM and HDM models. The attractive feature
of the
model of strings with HDM is the transfer of power to large scales without
completely erasing the power on small scales. This is in agreement with
observations
showing more power on large scales than predicted by the standard adiabatic+CDM
model
and also allows for the independent growth and formation of structures on
smaller
scales. Detailed N-body simulations of the string+HDM model attempting to
explore
in more detail these features are currently in progress.

\subsection{Gravitational Radiation}

 A particularly interesting cosmological constraint that can be imposed
on the cosmic string model is based on the gravitational radiation
produced by oscillating massive loops\cite{hr84,ca92}. These oscillations are
expected to lead
to a stochastic gravitational wave background which could be detectable as
perturbations of the period of msec pulsars\cite{srtr90}.
 The phase $\varphi (t)$ of the periodic signal detected by a msec pulsar
can be expanded as
\be
\varphi (t) = \varphi_0 + {\dot \varphi} t - {1\over 2} {\ddot \varphi} t^2
+ \varphi_R (t)
\ee
the first three terms can be modeled based on observations and pulsar physics
while $\varphi_R (t)$ is called {\it the residual phase} and is due to
perturbations of the signal period induced by either gravitational
waves or by other sources (pulsar intrinsic noise \etc). The quantity that
characterizes the energy density of gravitational waves of angular
frequency $\omega$ is
\be
\Omega_g (\omega)\equiv {{\omega \rho_g (\omega)}\over \rho_0}
\ee
where $\rho_g$ is the energy density in gravity waves.
For the stochastic background produced by string loops it may be shown that
\be
\Omega_g (\omega,G\mu) = 2.5 \times 10^{-8} \mu_6 ({{P}\over {2\pi}})^2
<\varphi_R^2 (T)> h^{-4} ({{2\pi}\over T})^4
\ee
where
\be
<\varphi_R^2 (T)> \equiv <{1\over T} \int_t^{t+T} \varphi_R^2 dt>
\ee
and P is the period of the signal in sec. The presently observed
time residual is $P <\varphi_R^2>/2\pi\simeq 1.5\mu sec$ and this implies
an upper bound for $\mu_6$ whose precise value depends on the
string simulations.
The most recent careful study of constraints from the msec pulsar\cite{ca92}
gives
a bound on $\mu_6$ of $\mu_6 \lsim 2\times 10^{-6} h^{-8/3}$. This result is
based on
the numerical simulations of Allen and Shellard\cite{as90} and it is projected
that
if the residual timing noise remains constant the bound in 1998 will be
 $\mu_6 \lsim 5\times 10^{-8} h^{-8/3}$ thus effectively ruling out the model.

\subsection{A String Detection?}

 The detection of cosmic string segments through the detection of multiple
lensing pairs of galaxies or quasars is a very exciting prospect.
 The expected properties of such lensing events can be summarized
as follows
\begin{itemize}
\item
A number of $N\gsim 3$ of galactic or quasar twin pairs is expected
concentrated in a small region in the sky (\eg $50'' \times 50''$)
\item
The angular separation of pairs is expected to be approximatelly
constant of $O(1'')$ as shown in Eq. (63).
\item
In contrast to most typical lensing events strings are expected
to induce no magnification of the images during lensing due
to the conical nature of their metric.
\item
The members of the lensed pair are expected to have very similar
properties (redshifts, spectra \etc) since they originate from the
same object.
\item
The redshift of the pair is expected to be relatively large
to increase the probability for a string being present along
the line of sight ($z\gsim 0.1$).
\end{itemize}
 An event that effectively fulfills all the above properties
was detected in 1987\cite{ch87} by Cowie and Hu who detected 4 twin pairs of
galaxies
in a $50'' \times 50''$ angular region in the sky with typical angular
separation $3''$, no magnification and very similar properties. These
properties are shown in Table 1  (from Ref. \cite{ch87})
while a contour plot of the twin pairs is shown in Fig. 11.
\newpage

{\bf Table 1:} Galaxy Properties (Ref. \cite{ch87})
\vspace{8cm}

{\bf Figure 11:} {\it A candidate string detection:} Four galactic twin pair in
a sky area
 of $50\times 50$ arcsec (Ref. \cite{ch87}).
\vspace{8cm}

 The strength of the candidates however was reduced by a later publication
which showed that the images of the pair members do not match
in a satisfactory way when images are studied in radio band\cite{hpth90}.
 Even though the issue is far from being resolved there
is currently no clear evidence that these events are induced by
a string lensing.

\section{Cosmic Strings and the Microwave Background}
\subsection{Angular Spectrum Basics}

 A pattern of CMB temperature fluctuations is characterized by two classes of
properties: properties of the {\it angular power spectrum} and {\it
statistical}
properties (probability distribution function \etc). I will give an
introduction of
the angular spectrum basics\cite{e89} and discuss the relevant predictions of
the cosmic string
model obtained by using a simple analytical approximation\cite{pcob,pstat}. The
predicted statistical
properties of the CMB fluctuations will be discussed also.

 Consider a photon scattered for last time on the last scattering surface at
$t_{rec}$ and reaching an observer with observational resolution $\Theta$
at time $t_0$. One of the main sources of CMB fluctuations is the
{\it Sachs-Wolfe} effect which is the result of gravitational potential
fluctuations
on the last scattering surface. Thus, an initially isotropic photon wavefront
has to
climb out of a potential $\delta \Psi$ whose depth varies with direction. The
resulting
fluctuations $({{\delta T}\over T})_{\lambda}$, smoothed on a scale $\lambda$
on the
last scattering surface are
\be
({{\delta T}\over T})_{\lambda} \sim \delta \Psi_{\lambda}
\ee
The gravitational potential due to a mass overdensity $\delta \rho$ on a scale
$\lambda$ is
\be
\delta \Psi_{\lambda} \sim {{G \delta M}\over \lambda} \sim G\delta
\rho_{\lambda}
\lambda^2 \sim G \lambda^{(3+n)/2} \lambda^2 \sim \lambda^{(1-n)/2}
\ee
where use was made of $\delta \rho_\lambda \sim (k^3 P(k))^{1/2} \sim
k^{(3+n)/2}
\sim \lambda^{-(3+n)/2}$. Now, Eq. (96) and Eq. (97) imply that
\be
({{\delta T}\over T})_{\Theta} \sim \delta \Phi_{\Theta}\sim \Theta^{(1-n)/2}
\ee
and therefore the correlation function at zero lag ($\Delta \theta =0$) for
angular resolution $\Theta$ is
\be
C(\Delta \theta =0)_\Theta \equiv <({{\delta T}\over T})^2>=
({{\delta T}\over T})_{rms}^2\sim \Theta^{(1-n)}
\ee
where $<>$ denotes ensemble average (equivalent to angular averaging with the
ergodic hypothesis). Therefore for a scale invariant power spectrum $(n=1)$,
$({{\delta
T}\over T})_{rms}$ is independent of the angular resolution
\be
({{\delta T}\over T})_{rms} \sim constant
\ee
Note that even for scale invariant primordial spectra (100) will be violated
on angular scales less than the angular scale of the horizon at recombination
($\Theta_{rec} \simeq z_{rec}^{-1/2}\simeq  0.03 rad \simeq 2^0$) due to
microphysical processes taking place on subhorizon scales (Doppler effect
\etc). Such
processes are discussed below.

CMB fluctuations are usually measured on large parts of the sky and in some
cases
over the whole celestial sphere. Therefore a convenient basis to analyse these
perturbations is not the Fourier basis but the {\it spherical harmonics}. A
fluctuation pattern
${{\delta T}\over T} ({\hat q})$ can be expanded as
\be
{{\delta T}\over T} ({\hat q}) = \sum_{l,m} a_l^m Y_l^m (\theta,\phi)
\ee
Using the addition theorem, the lack of preferred direction and defining $C_l
\equiv
<|a_l^m|^2>$ we may expand the angular correlation function $C(\theta)$ using
the
angular spectrum coefficients $C_l$
\be
C(\theta) \equiv  <{{\delta T}\over T} ({\hat q}_1) {{\delta T}\over T} ({\hat
q}_2)>= {1\over {4\pi}} \sum_{l=0}^\infty (2l+1) C_l P_l (\cos \theta)
\ee
where $\cos \theta \equiv {\hat q}_1 \cdot {\hat q}_2$.
 For observational experiments taking place on a relatively small part of the
sky the
contribution of low values of $l$ is filtered out and the above expression
reduces
as expected to a two dimensional Fourier transform with the angular spectrum
$C_l$
playing the role of the power spectrum $P(k)$
\ba
C(\theta)& = &{1\over {4\pi}} \sum_{l>>1}^\infty (2l+1) C_l P_l (\cos
\theta)\simeq {1\over {2\pi}} \sum_{l>>1} l C_l J_0 (l \theta) \Rightarrow \\
C(\theta)_{\Theta_0} & \simeq & {1\over {(2\pi)^2}} \int d^2 l C_l e^{i {\vec
l}\cdot
{\vec \theta}} W (l-l_0)
\ea
where $W(l-l_0)$ is a filter function centered on the maximum sensitivity
mode $l_0$ and filtering out modes that are undetectable due to either limited
sky
coverage or limited resolution.

Consider the power law ansatz $C_l \sim l^\alpha$. The angular scale
corresponding to
the mode $l$ may be approximated by $\Theta_l = {{\pi}\over l}$. Therefore,
using Eq. (104)
we can express $({{\delta T}\over T})_{rms}$ in terms of the resolution
$\Theta_0$
and comparing with Eq. (99) we can express $\alpha$ in terms of the power
spectrum index
$n$ \ie
\be
C(0)_{\Theta_0} = ({{\delta T}\over T})_{rms}^2 \sim l_0^2 C_{l_0} \sim
\Theta_0^{-\alpha - 2}
\ee
Now comparing Eq. (99) with Eq. (105) we obtain $\alpha =n-3$ and therefore
\be
C_l \sim l^{n-3}
\ee
which is approximatelly valid for $l>1$. Fig. 12 (from Ref. \cite{bcdes94})
shows the form of the
angular spectrum $l^2 C_l$ as
predicted by some models based on adiabatic fluctuations.

\vspace{8cm}
{\bf Figure 12:} The CMB angular spectrum as predicted by some typical models
based
on gaussian scale invariant fluctuations. Using appropriate filter functions
for each
experiment the
predicted $({{\delta T}\over T})_{rms}$ can be obtained.

 Notice the flatness of the
spectrum ($l^2 C_l =const$) on angular scales less than the horizon at
recombination
($l \sim 100$) which indicates the presence of scale invariance. COBE, with
angular
resolution $\Theta_0 \simeq 10^0$ and all sky coverage is sensitive to $l\lsim
35$ and
therefore the COBE data can provide both the normalization and the primordial
spectral
index $n$. Even though the different ways to perform the analysis of the data
lead to
slight variations in the value of $n=1$, the data are consistent with a scale
invariant spectrum $n=1$ and seem to favor a small positive tilt (blue
spectrum,
$n>1$) over a negative tilt (red spectrum $n<1$). Most models based on
inflation tend to
favor a small negative tilt of the spectrum\cite{ll92}.

Several experiments collect temperature data along one dimension in the sky for
example along a meridian. The power spectrum index can also be derived by using
the
data from such one dimensional experiments. In order to avoid confusion with
the two
dimensional angular spectrum $C_l$, I will denote the one dimensional angular
power spectrum
by $P(k)$ and the variable conjugate to the angle along the geodesic circle
under
consideration by $k$. This should not be confused with the density spectrum
where a similar notation is usually used. The correlation function
for a one dimensional pattern is the Fourier transform of $P(k)$
\be
C^{1d}(\Delta \theta) \simeq {1\over {2\pi}} \sum_k P(k) e^{ik\Delta \theta}
W(k-k_0)
\ee
where $W(k-k_0)$ is a filter function corresponding to the resolution of the
experiment. By isotropy $C^{1d}(\Delta \theta) = C^{2d} (\Delta \theta)$ and
also
$k\simeq l$ since they are both conjugate of the angular scale $\Delta \theta$.
Therefore from Eq. (104) and Eq. (107) we obtain
\be
l^2 C_l \simeq k P(k) \sim k^{n-1},  \hspace{1cm} ({{\delta T}\over T})_{rms} =
C^{1d} (0) = C^{2d} (0)
\ee
These results will be applied in approximating the predicted angular power
spectrum of the
string model.

\subsection{String Angular Spectrum}

There are three main sources that can lead to CMB temperature fluctuations in
the
context of the cosmic string model:
\begin{itemize}
\item
{Kaiser-Stebbins}\cite{ks84,g85} perturbations due to moving
long strings present between $t_{rec}$ and the present.
\item
Potential fluctuations on the last scattering surface ({\it Sachs-Wolfe}
effect\cite{sw67}) induced
by long string wakes and loops with their gravitationally accreted matter.
\item
{\it Doppler perturbations} induced by the local peculiar velocities of the
plasma on
which photons scatter for last time (see \eg Ref. \cite{p93}).
\end{itemize}
The above three types of perturbations are shown schematically in Fig. 8.
 Therefore the total string induced CMB temperature fluctuation may be wtitten
as
\be
({{\delta T}\over T})_{tot} = ({{\delta T}\over T})_{KS}+({{\delta T}\over
T})_{SW} +
({{\delta T}\over T})_{D}
\ee
Each one of these three contributions involves the superposition of some type
of
seeds and therefore in order to calculate it we must address  the following two
questions
\begin{enumerate}
\item
How can seeds be superposed?
\item
What type of seed should be superposed in each case?
\end{enumerate}
 I will here briefly address the first question. A detailed study of both
questions
may be found in Ref. \cite{pspec}. For simplicity I will focus on one
dimensional data.
Consider a geodesic circle in the sky and a seed CMB temperature
fluctuation function $f_1(\theta)$ with amplitude $a_1$ and angular scale
$\Psi$
(Fig. 13) centered at a random angular position $\theta_1$.

\vspace{6cm}
{\bf Figure 13:} Superposition of a seed on a geodesic circle on the sky.

 A random superposition of $N$ such seed functions will lead to a temperature
pattern of the form
\be
f(\theta)= \sum_{n=1}^{N} a_n f_1^\Psi (\theta-\theta_n) =
\sum_{n=1}^N a_n {1\over {2\pi}} \sum_{k=-\infty}^{+\infty} {\tilde f_1}^\Psi
(k)
e^{i k (\theta-\theta_n)}
\ee
and therefore the Fourier transform ${\tilde f} (k)$ of the pattern may be
written
as
\be
{\tilde f}(k) = {\tilde f_1}^\Psi (k) \sum_{n=1}^N a_n e^{i k \theta_n}
\ee
Thus, the power spectrum corresponding to the pattern is
\be
P_0 (k) = < |{\tilde f} (k)|^2> = N <|a_n|^2> |{\tilde f_1}^\Psi (k) |^2
\ee
In realistic cases however when the seed perturbations are induced by
topological
defects, the size $\Psi$ of the seed function will not be fixed but will grow
as
the horizon scale (the characteristic coherence scale of scaling defects)
grows.
Thus defect perturbations produced at later times when the horizon scale is
larger
will have a larger characteristic angular scale in the sky.
Assume for example that the comoving horizon grows by a scale factor $\alpha$
\ie $t^{com} \rightarrow \alpha t^{com}$. This implies that the angular scale
of
the horizon $\Theta_h$ and the seed size $\Psi$ will grow by the same factor
while
the total number of seeds $N$ superposed along a circle at this later time will
be
smaller by the same factor
\be
\Theta_h \rightarrow \alpha \Theta_h \Rightarrow (\Psi\rightarrow \alpha
\Psi,\hspace{0.5cm} N\rightarrow N/\alpha)
\ee
The power spectrum $P_Q (k)$ corresponding to the resulting pattern after $Q$
such
expansion steps may be written as
\be
P_Q (k) = \sum_{q=0}^Q {N\over {\alpha^q}} |f_1^{\alpha^q \Psi} (k)|^2
<|a_n|^2>
\ee
where the total number of steps $Q$ may be obtained from the ratio of the
maximum
over minimum seed (or horizon) size while the total number of seeds at the
first
expansion step is the number of defects per horizon scale (obtained from the
simulations leading to the scaling solution) times the total number of horizons
present on the circle during the first expansion step
\be
\Psi_{max} = \alpha^Q \Psi_{min} \hspace{1cm} N=M {{2\pi} \over \Theta_{min}}
\ee

Eq. (114) may now be used to find the contribution to the total spectrum from
each one of
the following perturbation types
\begin{itemize}
\item
Kaiser-Stebbins (KS) perturbations
\item
Potential perturbations due to wakes (W) and loops (L) with their accreted
matter
present on the last scattering surface.
\item
Doppler (D) perturbations due to plasma velocities induced by strings (photons
scattered
on moving plasma suffer Doppler shift).
\end{itemize}
The quantities that need to be specified in Eq. (114) for each one of the above
perturbation types are, the type of seed function $f_1$, the number of seeds
per
horizon $M$ and the range of seed scales that need to be superposed. Assuming
independent contributions from each perturbation type, the total spectrum
$P_{tot}$
may be obtained as a sum of the partial spectra as
\be
P_{tot} (k) = P_{KS} (k) + P_W (k) + P_L (k) +  P_D (k)
\ee
The detailed derivation of the form of $P_{tot}$ may be found in Ref.
\cite{pspec}. There it
is shown that $P_{tot} (k)$ depends on four parameters; the single free
parameter of
the cosmic string model $G\mu$ which may be normalized by comparing with the
CMB COBE
data or by large scale structure observations and three other parameters which
may in
principle be fixed by comparing with numerical simulations. These parameters
are
defined as
\ba
b &\equiv & M <(v_s \gamma_s)^2 > \\
f &\equiv & 1 + {{1-T/\mu_{eff}}\over {2 (v_s \gamma_s)^2}}\\
\xi &: &\Psi = \xi \Theta_h (t) /2
\ea
where $\xi$ is the string curvature radius $\Psi$ as a fraction of the horizon
scale $\Theta_h (t)$ and $f$ determines the wiggliness of long strings as
defined
previously. These three parameters may be fixed either by comparing directly
with
string simulations or by comparing the CMB spectral contribution $P_{KS} (k)$
of Eq. (116)
with the corresponding spectrum derived by propagating a photon wavefront
through a
simulated string network\cite{bsb92,bbs88}\footnote{$P_{tot}$ can not be used
because CMB simulations
with strings have not included so far the effects of potential and Doppler
fluctuations}.
Both of these approaches have been pursued in Ref. \cite{pspec} with results
that are
consistent with each other.
\newpage

{\bf Figure 14:} The total spectrum (a) including Kaiser-Stebbins, Sachs-Wolfe
and Doppler fluctuations obtained as discussed in the text. The contribution of
each individual component is also shown (b) (Ref. \cite{pspec}).
\vspace{8cm}

The total spectrum $P_{tot} (k)$ in units of $(G\mu)^2$, obtained after the
above
normalization, is shown in Fig. 14a while the contribution of each type of
perturbation is shown in Fig. 14b

Doppler CMB fluctuations are due mainly to long strings present on the last
scattering surface and therefore their characteristic scale corresponds to the
coherence scale of those strings. By the scaling solution, this scale is of the
order of the horizon at $t_{rec}$ ($\Theta_h (t_{rec}) \simeq 2^0 \Rightarrow
k_h
(t_{rec}) \simeq 100$). This explains the existence of a well defined peak for
the
Doppler term. The fact that the magnitude of this peak is significantly larger
than
the magnitude of the scale invariant Kaiser-Stebbins (KS) term may be
understood as
follows: The magnitude of the contribution to the KS term by each long string
is
\be
({{\delta T}\over T})_{KS} = 4\pi G\mu v_s \gamma_s {\hat
k} \cdot ({\hat v_s} \times {\hat s})
\ee
where ${\hat k}$ is the unit photon wave-vector and ${\hat s}$ is the unit
vector
along the string. The corresponding contribution to the Doppler term is
\be
({{\delta T}\over T})_D = {\hat k}\cdot {\delta {\vec v}} = 4\pi G\mu v_s
\gamma_s {\hat
k}\cdot ({\hat v_s} \times {\hat s}) f =({{\delta T}\over T})_{KS} f
\ee
 where $f\simeq 6$ ($f$ defined in Eq. (118)) according to
simulations\cite{bb88}. Thus we expect
the Doppler term to dominate over the KS term for $k\simeq k_h (t_{rec}) \simeq
100$
as is in fact seen in Fig. 14b
The negligible contribution of loops (dotted line in Fig. 14b)  to the total
spectrum
may also be understood by considering the fact that the typical loop radius is
a
tiny fraction (about $10^{-4}$) of the horizon as shown by simulations.
Therefore a
typical loop present on the last scattering surface at $t_{rec}$ corresponds to
an
angular scale of about $0.3$ arcsec which is way above the resolution of any
present
experiment.

It is now easy to normalize the remaining parameter $G\mu_{eff}$ by using the
COBE
DMR data. The angular correlation function predicted for the COBE (DMR)
experiment can be
expressed in terms of $P_{tot} (k)$ and the window function $W(k-k_0) \simeq
e^{-{{k^2}\over {(2\cdot 18)^2}}}$ as shown in Eq. (107). By demanding
agreement with the
detected $({{\delta T}\over T})_{rms}$ we have
\be
({{\delta T}\over T})^{DMR}_{rms} = (C(0))^{1/2} = 1.1 \times 10^{-6}
\Rightarrow (G\mu)_{eff} = 1.6 \times 10^{-6}
\ee

\vspace{8cm}
{\bf Figure 15:} The cosmic string predicted correlation function
smoothed on COBE scales. Superimposed are the first year COBE data (Ref.
\cite{pspec}).

This result is consistent with previous analytical studies\cite{pcob} and with
numerical simulation
studies\cite{bsb92,bbs88}. Fig. 15 shows $C(\theta)$ obtained by Eq. (107)
(normalized with Eq.
(122)) superposed with the COBE data. As expected by the scale invariant nature
of the string
perturbations on COBE scales, the agreement is fairly good.

The predicted power spectrum index can be found with a best fit based on Eq.
(108). By using
modes with $k\leq 20$ and the spectrum of Fig. 14a we find
\be
n \gsim 1.35 \pm 0.5
\ee
Eq. (123) is a lower limit to the actually predicted $n$ because Eq. (114)
gives a slight
overestimation of power on large scales\footnote{I thank Neil Turok for
pointing this out}
 by assuming
$\alpha$ fixed at all expansion steps (the expression $\Theta_h (t) \simeq
z(t)^{-1/2}$ was used for
all redshifts
$z(t)$ even though this approximation starts breaking down at low redshifts).

{\bf Table 2}:Detections of ${{\Delta T} \over T}_{rms}\times 10^{6}$ and the
corresponding
predictions of the string ($\Omega_0 =1$, $h=0.5$, no reionization,
$\Lambda=0$) and inflationary
models ($0.8\leq n
\leq 1.0$, $\Lambda=0$) normalized on COBE. \vskip 0.1cm
\begin{tabular}{|c|c|c|c|c|c|}\hline
Experiment & $k_0$ & $\Delta k$ & Detection & Strings & Inflation \\ \hline
COBE & 0 & 18 &11 $\pm$ 2&11 $\pm$ 3  & 11 $\pm$ 2 \\ \hline
TEN& 20 & 16 &  $\leq$  17 &13 $\pm$ 3  & 9 $\pm$ 1\\ \hline
SP91& 80 & 70 & 11 $\pm$ 5 &20 $\pm$ 5  & 12 $\pm$ 2 \\ \hline
SK&  85 & 60 & 14 $\pm$ 5 &19$\pm$ 4   & 12 $\pm$ 3 \\ \hline
MAX& 180 & 130 & $\leq$  30 ($\mu Peg$) &21 $\pm$ 5.5   &16 $\pm$ 5   \\ \hline
MAX& 180 & 130 & 49 $\pm$ 8 ($GUM$) &21 $\pm$ 5.5  & 16 $\pm$ 5  \\ \hline
MSAM &  300 & 200 & 16 $\pm$ 4   &19 $\pm$ 4    & 24 $\pm$ 6   \\ \hline
OVRO22 &  600 & 350 & - &13 $\pm$ 4    & 17$\pm$ 7  \\ \hline
WD&  550 & 400 &  $\leq$  12  & 17.5 $\pm$ 4.5    & 7 $\pm$2   \\ \hline
OVRO&  2000 & 1400 &  $\leq$  24 & 13.5 $\pm$ 3.5   & 7 $\pm$ 3  \\ \hline
\end{tabular}

 If the
exact $\Theta_h (z)$ relation was used, $\alpha$ would need to be larger at
late times
thus reducing the number of expansion steps at large scales. The result would
be a
slightly reduced power on large scales and a tilt of the spectrum towards $n>1$
for
the scale invariant KS contribution. This effect was taken into account in Ref.
\cite{bsb92}
where it was shown that it can increase the KS contribution to $n$ by about
$40$\%.
Here, the effect on the total spectrum will be smaller
since the other contributions remain unaffected by this.
Using $W(k)=e^{-(k-k_0)^2/{\Delta k}^2}$ and fixing $k_0$, $\Delta k$ for some
of the
ongoing experiments
we
are in position to  predict the corresponding value of ${{\Delta T} \over
T}_{rms}$ thus testing the
cosmic string model. These predictions with $1\sigma$ errors coming from the
variance of $a_n^2$ are
shown in Table 2. I also show some of the detections and upper limits existing
to date\cite{exper} as
well as the predictions of inflationary models for $0.8\leq n \leq 1.0$,
$\Lambda=0$
\cite{bcdes94}). At this time both inflationary models and cosmic strings
appear to be consistent
with detections at the $1\sigma$ level. However, as the quality of observations
improves, this may
very well change in the near future.

\subsection{Statistical Tests}

The predictions of the string model on the CMB angular spectrum are useful in
testing the model by
comparing the amplitude of the predicted CMB fluctuations at various angular
scales with the
corresponding observations. However, as shown in Table 1, these predictions can
not {\it
distinguish} the string model from models based on inflation. A potentially
interesting way that
can lead to this distinction between models is the study of the non-gaussian
character of string
induced fluctuations. Before discussing a simple model that can lead to the
identification of this
non-gaussian character I will give a brief introduction to some statistics
basics\cite{f71} that will
be used later.

A pattern of fluctuations of a random variable $\delta$ is characterized by the
probability
distribution $P(\delta)$ that gives the probability that a value $\delta$ will
be detected after a
random sampling of the pattern. The $n^{th}$ moment of the distribution
$P(\delta)$ is defined as
\be
<\delta^n> \equiv \int d\delta \hspace{3mm} \delta^n P(\delta)
\ee
The most useful moments are $\mu\equiv <\delta>$ (the mean), $\sigma^2
=<\delta^2>$ (the variance),
$a_3 \equiv {{<\delta^3>}\over {\sigma^3}}$ (the skewness) and $a_4 \equiv
{{<\delta^4>}\over
{\sigma^4}}$ (the kurtosis). A distribution is completely defined by its
infinite set of moments.

The moment generating function is a very useful function that can produce by
differentiation all the
moments of a distribution. It is defined as
\be
M_\delta (t) \equiv \int d\delta \hspace{3mm} e^{t\delta} P(\delta)
\ee
{}From this definition it immediatelly follows that
\be
<\delta^n> = {{d^n}\over {dt^n}} M(t)|_{t=0}
\ee
An interesting example is the gaussian distribution defined as
\be
P(\delta) = {1\over {\sqrt{2\pi \sigma^2}}} e^{-{{(\delta-\mu)^2}\over
{2\sigma^2}}} \rightarrow
P(\delta') = {1\over {\sqrt{2\pi}}} e^{-{{\delta'^2}\over {2}}}
\ee
where $\delta' \equiv {{\delta-\mu} \over \sigma}$ is a {\it standardized}
random variable.
The generating function corresponding to the standardized gaussian distribution
is obtained from
Eq. (127) and Eq. (125)
\be
M_{\delta'}(t) = e^{t^2/2} \Rightarrow a_3 = 0, \hspace{1cm} a_4 = 3
\ee
A particularly useful theorem that will be used below states that the moment
generating function of
a sum of two independent random variables is the product of the generating
functions
corresponding to each variable
\be
M_{\delta_1 + \delta_2} = M_{\delta_1} M_{\delta_2}
\ee

Another basic theorem of statistics is the {\it Central Limit Theorem}
(hereafter CLT) which states
that the sum
$\delta_n = x_1 + ... + x_n$ of a large number of independent random variables
$x_i$ with
identical distributions $P(x_i)$ has a probability distribution that approaches
the gaussian in the
limit of $n\rightarrow \infty$.

The above concepts apply in an interesting way to the CMB temperature
fluctuations.
The primordial fluctuations produced during inflation are due to the quantum
fluctuations of an
almost free scalar field. Therefore they may be represented as a Fourier series
where the phase of
each mode is random and uncorrelated with others \ie
\be
\delta = {{\delta T} \over T} (\theta) \sim \sum_k |\delta_k | e^{i \theta_k}
e^{i k x}
\ee
where $\theta_k$ is a random phase. Therefore, since $\delta$ is a sum of an
infinite number of
random variables, by the CLT it will obey gaussian statistics.

The CMB fluctuation pattern induced by cosmic strings may be expressed as a
superposition of seed
functions as in Eq. (110). In what follows I will focus on the KS term and find
the probability
distribution $P(\delta)$ of fluctuations induced by this
term\cite{pstat,mpb94,gp94,pfour}. The
result obtained in this way will only be an upper bound to the non-gaussian
character of fluctuations
because by the CLT any additional random effect can only make the fluctuations
more gaussian.

 Consider a step function perturbation with spatial size $2\Psi$ superposed at
a random position on a
pixel lattice with size $2L$ with periodic boundary conditions. The probability
that the
temperature of a pixel will be shifted positively (negatively) after this
superposition is $p={\Psi
\over L}$ ($p={\Psi\over L}$). Thus the moment generating function of the
fluctuation probability
distribution after this single superposition is
\be
M_{x_1} (t) = <e^{t x_1}> = (pe^t + pe^{-t} + (1-2p))
\ee
The corresponding result after the superposition of $n$ seed functions at
random positions is
\be
M_{\delta_n \equiv x_1 + ... + x_n} = (2p \cosh t + (1-2p))^n
\ee
where use of the theorem (129) was made. Assuming small $t$ (since to get the
moments we set $t=0$
after differentiation) and $p$ while taking large $n$ we may write
\be
M_{\delta_n}\simeq (1+ {{(2pn) t^2}\over {2n}})^n \rightarrow e^{\sigma t^2 /2}
\ee
where $\sigma^2 \equiv 2pn$ is the variance of the gaussian distribution
obtained (as expected by
the CLT) for a large  number $n$ of superposed seeds.

The moments of the distribution may now be easily obtained from Eq. (132) with
the help of Eq. (126)
as
\ba
\sigma^2 &=& {{d^2} \over {dt^2}} M_\delta (t)|_{t=0} = 2np \\
a_3 &=& {{d^3} \over {dt^3}} M_\delta (t)|_{t=0} = 0 \\
a_4 &=& {{d^4} \over {dt^4}} M_\delta (t)|_{t=0} = 3+ {{1-6p}\over {2np}}
\rightarrow 3
\ea
and the gaussian value of the kurtosis is approached as $1/n$.
Clearly the crucial quantity that determines how close we are to the gaussian
probability
distribution is the product $np$.  By setting $p={\Psi \over L}$ and $n= M
{L\over\Psi}$ (number of
seeds per horizon times the number of horizons in the lattice) we obtain $np =
M \simeq 10$. With
this value, the relative deviation of the kurtosis from the gaussian is
\be
{{a_4 -a_4^{gaussian}}\over a_4} \lsim {{1/2M}\over 3} \simeq {1\over 60} \ll 1
\ee
and any non-gaussian feature is undetectable. Can $np$ be decreased in order to
amplify the
non-gaussian features of the pattern? Clearly, in order to change $n$ we would
have to change
the model under study since $n$ is determined by the scaling solution whose
parameters are fixed
for cosmic strings. In order to change $p$ we would only have to change the
type of seed function.
This is indeed possible by considering the probability distribution of the
temperature of a linear
(or non-linear) combination of neighboring pixels. The most convenient
combination is the
temperature difference of neighboring pixels since this leads (for step-like
temperature seeds)
 to a
very localized seed function of
$\delta$-function type. The number of affected pixels (and therefore $p$) is
minimized by such a
seed function. Let us therefore calculate the moment generating function
corresponding to the
variable $d\equiv \delta^{i+1} - \delta^{i}$ where the superscript now denotes
pixel location. By
defining a new probability $q\equiv {1\over L}$ that a pixel will be affected
by a given seed we
have (in analogy with Eq. (132))
\be
M_{x_1}= <e^{tx_1}> = (2qe^t + 2 q e^{-t} + q e^{2t} + qe^{-2t} + (1-6q))
\ee
and following the same steps as above we can obtain the moments for the new
variable $d$ as
$a_3 = 0$ (due to the symmetry of the seed function) and
\be
a_4 = 3 + {{1-12q} \over {4nq}}
\ee
But now $q={1\over L}$ and $n=M {L\over \Psi}$ which implies
\be
nq\sim {M\over \Psi} \ll M =np
\ee
and the relative deviation of the kurtosis from the gaussian is
\be
{{a_4 -a_4^{gaussian}}\over a_4} \simeq {{\Psi/4M}\over 3}= {\Psi \over {12 M}}
\gg {1\over {6M}}
\ee
 The above described test based on the CMB temperature gradient is effective if
the the number of
pixels $\Psi$ affected by half the step function is much larger than 2 (see Eq.
(141)). Therefore
we require
\be
\Psi \equiv {{\Psi^0} \over {\delta \Theta_{pix}^0}}>
{{\Psi^0_{min}} \over {\delta \Theta_{pix}^0}} \simeq {{2^0} \over {\delta
\Theta_{pix}^0}}
\gg 2
\ee
where $\delta \Theta_{pix}^0$ is the pixel size in degrees.
Thus the statistics of the temperature gradient patterns provide a more
sensitive test for
non-gaussianity than statistics of temperature patterns\cite{gpjbbbs90,mpb94}
in experiments with
resolution of  ${\delta \Theta_{pix}^0}\simeq 1^0$ or better.

 A more detailed analytical study of the moments of the string
induced CMB fluctuations taking into account the growth
of the seed functions as the horizon expands may be found in
Refs. \cite{mpb94,pstat} Those analytical results can be tested by using
Monte Carlo simulations performed as follows:
 A one dimensional array of 512 initially unperturbed pixels is
considered and a number $N=M {{512}\over {\Psi_{min}}}$
of step-like seeds ($\Psi_{min}$ is
the initial seed size in units of pixel size which is determined
by the resolution of the simulated experiment) is superposed on
the array at random positions and with periodic boundary conditions.
 To simulate the growth of the horizon a larger seed size
$\Psi = \alpha \Psi_{min}$ is considered next and the superosition
is repeated for $Q$ such steps until the seed size becomes much
larger than the size of the array of pixels (larger seeds shift the
whole lattice by a constant and do not affect the statistics).
 The resulting pattern is then Fourier transformed and gaussianized
by assigning random phases to the Fourier modes and reconstructing the
pattern. The gaussian and the stringy patterns are then compared
using different statistical tests, various resolutions and adding
random gaussian noise of various signal to noise ratios.
 The Monte Carlo pattern constructed along the above lines may be
written as
\be
{{\delta T} \over T} (\theta)|_{str} = \sum_{q=1}^Q \sum_{i=1}^{N/\alpha^q}
a_n f_1^{\alpha^q \Psi} (\theta - \theta_i)
\ee
while the Fourier transform is obtained as
\be
{\tilde {{\delta T} \over T}} (k) = {1\over {2L}} \int_{-L}^{L} d\theta e^{i k
\pi \theta / L}
{{\delta T} \over T} (\theta)|_{str}
\ee
and the gaussianized pattern with the same scale invariant power
spectrum is
\be
{{\delta T} \over T} (\theta)|_{gauss} = \sum_{k=-\infty}^{+\infty} |{\tilde
{{\delta T} \over T}}
(k)| e^{i\theta_k}  e^{i k \pi
\theta / L}
\ee
where $\theta_k$ is a random phase.
\newpage

{\bf Figure 16:} An exaggerated stringy temperture CMB pattern composed of
temperature Kaiser-Stebbins discontinuities(a) compared with its random phase
realization (b)
(Ref. \cite{gp94}).
\vspace{8cm}

{\bf Figure 17:} Plot of kurtosis of temperature fluctuations versus
angular resolution ($\theta_{min}$) with zero noise.  Even at the
lowest resolutions ($\sim 0.5 arcmin$), this test cannot distinguish the
gaussian from the
string pattern (Ref. \cite{gp94}).
\vspace{8cm}

Fig. 16a demonstrates a stringy pattern with unrealistically amplified
non-gaussian
features (small number of seeds, fixed seed size) while Fig. 16b
is the corresponding gaussianised (random phase pattern). These two
types of patterns may now be compared using statistical tests
based on the moments of the distributions.
After constructing 100
realizations in each case, the mean and standard deviation (cosmic variance
of string perturbations\cite{pvari}) of
the kurtosis is found for various experimental resolutions
starting from 0.5 arcmin to one degree. Even without noise
added this test proves inadequate to identify the non-gaussian
features of the stringy pattern (Fig. 17).

This could have been expected in view
of the above analytical considerations. The corresponding
test for the temperature difference of neighboring pixels however
is much more efficient as shown in Fig. 18a where the kurtosis
of the two types of patterns are compared. For resolutions better
than a few arcminutes the non-gaussian features of the stringy pattern
are manifest through a significant increase of the kurtosis.

\vspace{8cm}
{\bf Figure 18:} Plot of kurtosis of temperature gradient of
fluctuations versus angular resolution.  For small resolutions and
zero noise (a), this test can distinguish between seed and gaussian
patterns. With gaussian instrumental noise
added (b) noise to signal ratio of $n/s = 0.3$ destroys all trace of the
non-gaussian features of the seed pattern. The signal survives at $n/s \lsim
0.1$
(Ref. \cite{gp94}).

Unfortunately
this nice feature of this test is very sensitive to the addition
of noise to the fluctuation pattern as shown in Fig. 18b  which
shows the effect of adding a modest amount of noise (the
noise to signal ratio was $n/s=0.3$) on the kurtosis
of the temperature difference pattern.
This sensitivity to
noise of the temperature difference patterns is due to the
low temperature derivative for most pixels in temperature
patterns constructed by step-function superposition compared
to the much larger derivative of noise patterns. Thus it is
much easier for noise to dominate in temperature difference (derivative)
patterns.

\section{Conclusion}

  The purpose of this review was to give an introduction to the basic
features of the cosmic string model for large scale structure formation
and to show some of the recent developments in testing the model by comparing
with new detailed observational data. The main points stressed are the
following:
\begin{enumerate}
\item
Cosmic strings are linear concentrations of trapped energy predicted by
GUTs to form during a phase transition in the early universe.
\item
For a natural value of a single free parameter $\mu$ (mass per unit length)
 strings may
provide the seeds for structure formation.
\item
The main achievements of the model include the following
\begin{itemize}
\item
Three independent types of requirements lead to the same value
for the free parameter $\mu$ of the model: {\it Large Scale Structure,
Microwave Background, Grand Unified Theories}.
\item
There is a well defined mechanism for the formation of
the observed sheets and filaments of galaxies on large scales.
\item
The model can resurrect massive neutrinos as dark matter candidates
since strings+HDM appears to be a viable model for structure
formation with amplified power on large scales
(seeds survive free streaming).
\item
The statistics and power spectrum of the predicted CMB fluctuations
are consistent with current CMB observational data.
\item
The existence of a scaling solution makes the model consistent with
standard cosmology in contrast with other models based on topological defects.
\end{itemize}
\item
The main challenges of the model include the following
\begin{itemize}
\item
The complexity of performing detailed simulations of structure
formation including both the string network and matter. This
results to a large uncertainty in the predictions of the model.
\item
The $L^{-1}$ scaling of the predicted velocity fields with scale $L$
is hard to reconcile with large scale velocity field observations.
\item
The msec pulsar constraints appear to uncomfortably close to ruling
out the model.
\end{itemize}
The model makes some well defined unique predictions that can be used
to rule out or verify it
\begin{itemize}
\item
The CMB fluctuations are predicted to have certain non-gaussian features
on scales of $1'$ or smaller.
\item
The model predicts the concentration of multiple images of lensed
objects in small areas in the sky with certain unique characteristics.
Such an event may have already been detected.
\end{itemize}
\end{enumerate}

\bigskip

{\bf Acknowledgements}

\noindent
It is a pleasure to thank Qaisar Shafi and all the organizers of the Trieste
Summer School for an invitation to lecture at an exciting and well organized
school. This work was supported by a CfA Postdoctoral fellowship
and by DOE Grant DE-FG0291ER40688, Task A.


\vskip 1cm

\begin{thebibliography}{99}
\bibitem{w72}see e.g., S. Weinberg, `Gravitation and Cosmology'
(Wiley, New York, 1972).
\bibitem{m33}E. Milne, {\it Zeits. f. Astrophys.} {\bf 6}, 1 (1933).
\bibitem{m90}J. Mather et al., {\it Ap. J. (Lett.)} {\bf 354}, L37
(1990).
\bibitem{h27}E. Hubble, {\it Proc. Nat. Acad. Sci.} {\bf 15}, 168
(1927).
\bibitem{p88}R. Partridge, {\it Rep. Prog. Phys.} {\bf 51}, 647 (1988).
\bibitem{pw65}A. Penzias and R. Wilson, {\it Ap. J.} {\bf 142}, 419
(1965).
\bibitem{fcw83}D. Fixen, E. Cheng and D. Wilkinson, {\it Phys. Rev. Lett.} {\bf
50}, 620 (1983).
\bibitem{ah50}R. Alpher and R. Herman, {\it Rev. Mod. Phys.} {\bf
22}, 153 (1950)\\
G. Gamov, {\it Phys. Rev.} {\bf 70}, 572 (1946).
\bibitem{dprw65}R. Dicke, P.J.E. Peebles, P. Roll and D. Wilkinson,
{\it Ap. J.} {\bf 142}, 414 (1965).
\bibitem{abg48}R. Alpher, H. Bethe and G. Gamov, {\it Phys. Rev.}
{\bf 73}, 803 (1948)\\
R. Alpher and R. Herman, {\it Nature} {\bf 162}, 774 (1948).
\bibitem{ssotkl92}S. Shechtman, P. Schechter, A. Oemler, D. Tucker, R. Kirshner
and
H. Lin, Harvard-Smithsonian preprint CFA 3385 (1992), to appear in `Clusters
and
Superclusters of Galaxies', ed. by A. Fabian (Kluwer, Dordrecht, 1993)\\
S. Shechtman et al., `The Las Campanas Fiber-Optic Redshift Survey',
CFA preprint (1994),
to be publ. in the proc. of the 35th Herstmonceaux Conference
"Wide Field Spectroscopy and the Distant Universe".
\bibitem{zes82}Ya.B. Zel'dovich, J. Einasto and S. Shandarin, {\it Nature}
{\bf 300}, 407 (1982)\\
J. Oort, {\it Ann. Rev. Astron. Astrophys.} {\bf 21}, 373 (1983)\\
R.B. Tully, {\it Ap. J.} {\bf 257}, 389 (1982)\\
S. Gregory, L. Thomson and W. Tifft, {\it Ap. J.} {\bf 243}, 411
(1980).
\bibitem{bs83}N. Bahcall and R. Soneira, {\it Ap. J.} {\bf 270}, 20 (1983)\\
A. Klypin and A. Kopylov, {\it Sov. Astr. Lett.} {\bf 9}, 41 (1983).
\bibitem{lgh86}V. de Lapparent, M. Geller and J. Huchra, {\it Ap. J.
(Lett)} {\bf 302}, L1 (1986).
\bibitem{dh82}M. Davis and J. Huchra, {\it Ap. J.} {\bf 254}, 437
(1982).
\bibitem{t87}V. Trimble, {\it Ann. Rev. Astr. Astrophys.} {\bf 25},
423 (1987).
\bibitem{bd89}E. Bertschinger and A. Dekel, {\it Ap. J.} {\bf 336},
L5 (1989).
\bibitem{sdyh92}M. Strauss, M. Davis, A. Yahil and J. Huchra, {\it
Ap. J.} {\bf 385}, 421 (1992).
\bibitem{pss88}J. Primack, D. Seckel and B. Sadoulet, {\it Ann. Rev. Nucl.
Part. Sci.}
{\bf 38}, 751 (1988).
\bibitem{pww83} J. Preskill, M. B. Wise and F. Wilczek, {\it Phys. Lett.} {\bf
B120}, 127 (1983).
\bibitem{bes80}J. Bond, G. Efstathiou and J. Silk, {\it Phys. Rev. Lett.} {\bf
45},(1980)\\
J. Bond and A. Szalay, {\it Ap. J.} {\bf 274}, 443 (1983).
\bibitem{p80}W. Press, {\it Phys. Scr.} {\bf 21}, 702 (1980).
\bibitem{cm80}G. Chibisov and V. Mukhanov, `Galaxy Formation and
Phonons,' Lebedev Physical Institute Preprint No. 162 (1980)\\
G. Chibisov and V. Mukhanov, {\it Mon. Not. R. Astron. Soc.} {\bf
200}, 535 (1982).
\bibitem{l80}V. Lukash, {\it Pis'ma Zh. Eksp. Teor. Fiz.} {\bf 31}, 631
(1980).
\bibitem{s81}K. Sato, {\it Mon. Not. R. Astron. Soc.} {\bf 195},
467 (1981).
\bibitem{g81}A. Guth, {\it Phys. Rev.} {\bf D23}, 347 (1981).
\bibitem{k76} T.W.B. Kibble, {\it J. Phys.} {\bf A9}, 1387 (1976).
\bibitem{k80}T.W.G. Kibble, {\it Phys. Rep.} {\bf 67}, 183 (1980).
\bibitem{v85}A. Vilenkin, {\it Phys. Rep.} {\bf 121}, 263 (1985).
\bibitem{t89}N. Turok, `Phase Transitions as the Origin of Large-Scale
Structure,' in `Particles, Strings and Supernovae' (TASI-88) ed. by
A. Jevicki and C.-I. Tan (World Scientific, Singapore, 1989).
\bibitem{vs94}A. Vilenkin and E.P.S. Shellard, `Strings and Other Topological
Defects'
(Cambridge Univ. Press, Cambridge, 1994).
\bibitem{r85}see e.g., G. Ross, Grand Unified Theories (Benjamin, Reading,
1985).
\bibitem{bfpr84}G. Blumenthal, S. Faber, J. Primack and M. Rees, {\it Nature}
{\bf 311}, 517 (1984)\\
M. Davis, G. Efstathiou, C. Frenk and S. White, {\it Ap. J.} {\bf 292}, 371
(1985).
\bibitem{z80}Ya.B. Zel'dovich, {\it Mon. Not. R. astron. Soc.} {\bf 192},
663 (1980).
\bibitem{v81}A. Vilenkin, {\it Phys. Rev. Lett.} {\bf 46}, 1169 (1981).
\bibitem{s51} N. Steenrod, `The topology of fibre bundles' Princeton Univ.
Press. (1951).
\bibitem{pexis}L. Perivolaropoulos {\it Phys. Lett.} {\bf B316}, 528 (1993) and
references therein.
\bibitem{ponth}T. Vachaspati, {\it Nucl.Phys.} {\bf B397} (1993).\\
 M. James L. Perivolaropoulos and T. Vachaspati,
{\it Phys. Rev.} {\bf D46}, R5232 (1992).\\
M. James L. Perivolaropoulos and T. Vachaspati,
Bulletin Board: hep-ph@xxx.lanl.gov - 9212301,
{\it Nucl. Phys.} {\bf B395}, 534
(1993).\\
 A. Achucarro, K. Kuijken L. Perivolaropoulos and T. Vachaspati,
{\it Nucl. Phys.} {\bf B388}, 435 (1992).
\bibitem{m79}D. Mermin, {\it Rev. Mod. Phys.} {\bf 51}, 591 (1979).
\bibitem{g74}P. de Gennes, `The Physics of Liquid Crystals' (Clarendon Press,
Oxford, 1974)\\
I. Chuang, R. Durrer, N. Turok and B. Yurke, {\it Science} {\bf 251}, 1336
(1991)\\
M. Bowick, L. Chandar, E. Schiff and A. Srivastava, {\it Science} {\bf 263},
943 (1994).
\bibitem{sv87}M. Salomaa and G. Volovik, {\it Rev. Mod. Phys.} {\bf 59}, 533
(1987).
\bibitem{a57}A. Abrikosov, {\it JETP} {\bf 5}, 1174 (1957).
\bibitem{d87}R. Davis, {\it Phys. Rev.} {\bf D35}, 3705 (1987).
\bibitem{t89a}N. Turok, {\it Phys. Rev. Lett.} {\bf 63}, 2625 (1989).
\bibitem{d64} G. H. Derrick, {\it J. Math. Phys.} {\bf 5}, 1252 (1964).
\bibitem{r87}see e.g. R. Rajaraman, `{\it Solitons and Instantons}', North
Holland Publishing (1987), p. 78, and references therein.
\bibitem{p92}L. Perivolaropoulos, {\it Phys. Rev.} {\bf D46}, 1858
(1992).
\bibitem{psb92}T. Prokopec, A. Sornborger and R. Brandenberger, {\it Phys.
Rev.} {\bf D45}, 1971 (1992).
\bibitem{bcl91}J. Borrill, E. Copeland and A. Liddle, {\it Phys. Lett.} {\bf
258B}, 310 (1991).
\bibitem{s93}A. Sornborger, {\it Phys. Rev.} {\bf D48}, 3517 (1993).
\bibitem{ts90}N. Turok and D. Spergel, {\it Phys. Rev. Lett.} {\bf 64}, 2736
(1990).
\bibitem{gst91}A. Gooding, D. Spergel and N. Turok, {\it Ap. J. (Lett.)}
{\bf 372}, L5 (1991)\\
C. Park, D. Spergel and N. Turok, {\it Ap. J. (Lett.)} {\bf 373}, L53 (1991).
\bibitem{stpr91}D. Spergel, N. Turok, W. Press and B. Ryden, {\it Phys. Rev.}
{\bf D43}, 1038 (1991).
\bibitem{cost91}R. Cen, J. Ostriker, D. Spergel
and N. Turok, {\it Ap. J.} {\bf 383}, 1 (1991).
\bibitem{t91}N. Turok, {\it Phys. Scripta} {\bf T36}, 135 (1991).
\bibitem{no73}H. Nielsen and P. Olesen, {\it Nucl. Phys.} {\bf B61}, 45
(1973).
\bibitem{bv89}M. Barriola and A. Vilenkin, {\it Phys. Rev. Lett.} {\bf 63},
341 (1989).
\bibitem{rb90}S. Rhie and D. Bennett, {\it Phys. Rev. Lett.} {\bf 65}, 1709
(1990).
\bibitem{plarg}L. Perivolaropoulos,
Bulletin Board: hep-ph@xxx.lanl.gov - 9207255,
 {\it Mod. Phys. Lett.} {\bf A7} 903 (1992).
\bibitem{pasym}L. Perivolaropoulos,
Bulletin Board: hep-ph@xxx.lanl.gov - 9310264,
 {\it Phys. Rev.} {\bf D48}, 5961 (1993).
\bibitem{pscat} W. Perkins, L. Perivolaropoulos, R. Brandenberger, A. C.Davis
and A. Matheson,
{\it Nucl.
Phys.} {\bf B353} 237 (1991) and references therein.\\
R. Brandenberger and L. Perivolaropoulos, {\it Phys. Lett.} {\bf 208B} 396
(1988).\\
 A. Matheson, R. Brandenberger, A. C.Davis,and L. Perivolaropoulos and W.
Perkins,
{\it Phys. Lett.} {\bf 248B} 263 (1990).
\bibitem{vv84}T. Vachaspati and A. Vilenkin, {\it Phys. Rev.} {\bf D30},
2036 (1984).
\bibitem{f74}D. Foerster, {\it Nucl. Phys.} {\bf B81}, 84 (1974).
\bibitem{t88}N. Turok, in `Proceedings of the 1987 CERN/ESO Winter School
on Cosmology and Particle Physics' (World Scientific, Singapore,
1988).
\bibitem{s87}E.P.S. Shellard, {\it Nucl. Phys.} {\bf B283}, 624 (1987).
\bibitem{m88}R. Matzner, {\it Computers in Physics} {\bf 1}, 51 (1988)
K. Moriarty, E. Myers and C. Rebbi, {\it Phys. Lett.} {\bf 207B}, 411
(1988)\\
E.P.S. Shellard and P. Ruback, {\it Phys. Lett.} {\bf 209B}, 262
(1988).
\bibitem{pinst} L. Perivolaropoulos {\it Nucl. Phys.} {\bf B375} 655 (1992).
\bibitem{r88}P. Ruback, {\it Nucl. Phys.} {\bf B296}, 669 (1988).
\bibitem{vv85}T. Vachaspati and A. Vilenkin, {\it Phys. Rev.} {\bf D31},
3052 (1985)\\
N. Turok, {\it Nucl. Phys.} {\bf B242}, 520 (1984)\\
C. Burden, {\it Phys. Lett.} {\bf 164B}, 277 (1985).
\bibitem{at85}A. Albrecht and N. Turok, {\it Phys. Rev. Lett.} {\bf 54},
1868 (1985).
\bibitem{bb88}D. Bennett and F. Bouchet, {\it Phys. Rev. Lett.} {\bf 60},
257 (1988).
\bibitem{as90}B. Allen and E.P.S. Shellard, {\it Phys. Rev. Lett.} {\bf
64}, 119 (1990).
\bibitem{at89}A. Albrecht and N. Turok, {\it Phys. Rev. } {\bf D40}, 973
(1989).
\bibitem{c90}B. Carter, {\it Phys. Rev.} {\bf D41}, 3869 (1990).
\bibitem{cka92}E. Copeland, T.W.B. Kibble and D. Austin, {\it Phys. Rev.}
{\bf D45}, 1000 (1992).
\bibitem{v92}D. Vollick, {\it Phys. Rev.} {\bf D45}, 1884 (1992)
\bibitem{vv91} T. Vachaspati and A. Vilenkin, {\it Phys. Rev. Lett.} {\bf 67},
1057 (1991).
\bibitem{zk78}Ya.B. Zel'dovich and M. Khlopov, {\it Phys. Lett.} {\bf 79B},
239 (1978)\\
J. Preskill, {\it Phys. Rev. Lett.} {\bf 43}, 1365 (1979).
\bibitem{lp80}P. Langacker and S.-Y. Pi, {\it Phys. Rev. Lett.} {\bf 45}, 1
(1980).
\bibitem{zk74}Ya.B. Zel'dovich, I. Kobzarev and L.
Okun, {\it Zh. Eksp. Teor. Fiz.} {\bf 67}, 3 (1974).
\bibitem{v81a}A. Vilenkin, {\it Phys. Rev.} {\bf D23}, 852 (1981)\\
J. Gott, {\it Ap. J.} {\bf 288}, 422 (1985)\\
W. Hiscock, {\it Phys. Rev.} {\bf D31}, 3288 (1985)\\
B. Linet, {\it Gen. Rel. Grav.} {\bf 17}, 1109 (1985)\\
D. Garfinkle, {\it Phys. Rev.} {\bf D32}, 1323 (1985)\\
R. Gregory, {\it Phys. Rev. Lett.} {\bf 59}, 740 (1987).
\bibitem{ks84} N. Kaiser and A. Stebbins, {\it Nature} {\bf 310}, 391
(1984)\\
A. Stebbins, {\it Astrophys. J.} {\bf 327}, 584 (1988).
\bibitem{g85} R. Gott 1985, Ap. J. {\bf 288}, 422.
\bibitem{hn84} C. Hogan and R. Narayan, {\it M.N.R.A.S.}, {\bf 211} 575 (1984).
\bibitem{t84}N. Turok, {\it Nucl. Phys.} {\bf B242}, 520 (1984).
\bibitem{v86}T. Vachaspati, {\it Phys. Rev. Lett.} {\bf 57}, 1655 (1986).
\bibitem{sv84}J. Silk and V. Vilenkin, {\it Phys. Rev. Lett.} {\bf 53},
1700 (1984).
\bibitem{svbst87}A. Stebbins, S. Veeraraghavan, R.
Brandenberger, J. Silk and N. Turok, {\it Ap. J.}
{\bf 322}, 1 (1987).
\bibitem{bps90}R. Brandenberger, L. Perivolaropoulos and A. Stebbins, {\it
Int. J. of Mod. Phys.} {\bf A5}, 1633 (1990)\\
L. Perivolaropoulos, R. Brandenberger and A. Stebbins, {\it Phys.
Rev.} {\bf D41}, 1764 (1990)\\
R. Brandenberger, {\it Phys. Scripta} {\bf T36}, 114 (1991).
\bibitem{hm93} T. Hara  and S. Miyoshi,  {\it Ap. J.}
{\bf 405}, 419 (1993).
\bibitem{tb86}N. Turok and R. Brandenberger, {\it Phys. Rev.} {\bf D33},
2175 (1986)\\
A. Stebbins, {\it Ap. J. (Lett.)} {\bf 303}, L21 (1986)\\
H. Sato, {\it Prog. Theor. Phys.} {\bf 75}, 1342 (1986).
\bibitem{wfd83}S. White, C. Frenk and M. Davis, {\it Ap. J. (Lett.)}
{\bf 274}, L1 (1983).
\bibitem{bkst87}R. Brandenberger, N. Kaiser, D. Schramm and N. Turok, {\it
Phys. Rev. Lett.} {\bf 59}, 2371 (1987).
\bibitem{bkt87}R. Brandenberger, N. Kaiser and N. Turok, {\it Phys. Rev.}
{\bf D36}, 2242 (1987).
\bibitem{v92a} T. Vachaspati {\it Phys.Lett.} {\bf B282} 305, (1992).\\
L. Perivolaropoulos and T. Vachaspati,
Bulletin Board: hep-ph@xxx.lanl.gov - 9303242,
{\it Ap. J. Lett.} {\bf 423 }, L77 (1994).
\bibitem{h70}E. Harrison, {\it Phys. Rev.} {\bf D1}, 2726 (1970)
Ya.B. Zel'dovich, {\it Mon. Not. R. astron. Soc.} {\bf 160}, 1p
(1972).
\bibitem{gp82}A. Guth and S.-Y. Pi, {\it Phys. Rev. Lett.} {\bf 49}, 1110
(1982)\\
S. Hawking, {\it Phys. Lett.} {\bf 115B}, 295 (1982)\\
A. Starobinsky, {\it Phys. Lett.} {\bf 117B}, 175 (1982)\\
R. Brandenberger and R. Kahn, {\it Phys. Rev.} {\bf D28}, 2172 (1984)\\
J. Frieman and M. Turner, {\it Phys. Rev.} {\bf D30}, 265 (1984)\\
V. Mukhanov, {\it JETP Lett.} {\bf 41}, 493 (1985).
\bibitem{as92a}A. Albrecht and A. Stebbins, {\it Phys. Rev. Lett.} {\bf
68}, 2121 (1992).
\bibitem{as92b}A. Albrecht and A. Stebbins, {\it Phys. Rev. Lett.} {\bf 69},
2615 (1992).
\bibitem{hr84} C.J. Hogan and M.J. Rees, {\it Nature} {\bf 311}, 109 (1984).\\
B. Bertoti, B.J. Carr and M.J. Rees, {\it M.N.R.A.S.} {\bf 203}, 945 (1983).\\
F.R. Bouchet, D.P. Bennett, {\it Phys. Rev.} {\bf D41}, 720 (1990).
\bibitem{ca92} R.R. Caldwell and B. Allen, {\it Phys.Rev.} {\bf D45} 3447,
(1992).
\bibitem{srtr90}D.R. Stinebring, M.F. Ryba, J.H. Taylor (Princeton U.) and R.W.
Romani, {\it Phys.Rev.Lett.} {\bf 65} 285 (1990).
\bibitem{ch87} L. L. Cowie and E. M. Hu, {\it Ap. J.} {\bf 318} L33 (1987).\\
E.M. Hu, {\it Ap. J.} {\bf 360}, L7 (1990).
\bibitem{hpth90} J.N. Hewitt, R.A. Perley, E.L. Turner and E.M. Hu, {\it Ap.
J.} {\bf 356}, 57 (1990).
\bibitem{e89}see e.g., G. Efstathiou, in `Physics of the Early
Universe,' proc. of the 1989 Scottish Univ. Summer School in Physics,
ed. by J. Peacock, A. Heavens and A. Davies (SUSSP Publ., Edinburgh,
1990).
\bibitem{pcob}L. Perivolaropoulos, {\it Phys. Lett.} {\bf 298B}, 305 (1993).
\bibitem{pstat}L. Perivolaropoulos,
Bulletin Board: hep-ph@xxx.lanl.gov - 9403298,
  {\it Phys. Rev.} {\bf D50} 962 (1994).
\bibitem{ll92}A. R. Liddle, D. H. Lyth,
{\it The Spectral Slope in the Cold Dark Matter Cosmogony},
Bulletin Board: astro-ph@babbage.sissa.it - 9209003, SUSSEX-AST-92-8-1, (1992).
\bibitem{sw67}R. Sachs and A. Wolfe, {\it Ap. J.} {\bf 147}, 73 (1967).
\bibitem{p93}T. Padmanabhan, `Structure Formation in the Universe'
(Cambridge Univ. Press 1993).
\bibitem{pspec} L. Perivolaropoulos,
{\it Spectral Analysis of Microwave Background Perturbations Induced by Cosmic
Strings},
Bulletin Board: astro-ph@babbage.sissa.it - 9402024,
CfA-3796, submitted to {\it Ap. J.} (1994).
\bibitem{bsb92}D. Bennett, A. Stebbins and F. Bouchet, {\it Ap. J. (Lett.)}
{\bf 399}, L5 (1992).
\bibitem{bbs88} D. Bennett, F. Bouchet and A. Stebbins, {\it Nature} {\bf
335} 410 (1988).
\bibitem{exper}COBE: G. Smoot et al., {\it Ap. J. (Lett.)} {\bf 396}, L1
(1992).\\
MSAM: S. Meyer, E. Cheng and L. Page, {\it Ap. J. (Lett.)} {\bf 371}, L7
(1991)\\
MSAM: K. Ganga, E. Cheng, S. Meyer and L. Page, {\it Ap. J. (Lett.)} {\bf 410},
L57
 (1993).\\
S. Hancock et al., {\it Nature} {\bf 317}, 333 (1994).\\
SK: E. Wollack et al., {\it Ap. J. (Lett.)} {\bf 419}, L49 (1993).\\
SP91: T. Gaier et al., {\it Ap. J. (Lett.)} {\bf 398}, L1 (1992)\\
SP91: J. Schuster et al., {\it Ap. J. (Lett.)} {\bf 412}, L47 (1993).\\
P. de Bernardis et al., {\it Ap. J. (Lett.)} {\bf 422}, L33 (1994).\\
M. Dragovan et al., Princeton preprint (1993).\\
MAX: P. Meinhold et al., {\it Ap. J. (Lett.)} {\bf 409}, L1 (1993).\\
MAX: J. Gunderson et al., {\it Ap. J. (Lett.)} {\bf 413}, L1 (1993).\\
E. Cheng et al., {\it Ap. J. (Lett.)} {\bf 422}, L37 (1994).\\
WD: G. Tucker, G. Griffin, H. Nguyen and J. Peterson, {\it Ap. J. (Lett.)}
{\bf 419}, L45 (1993).\\
OVRO: A. Readhead et al., {\it Ap. J.} {\bf 346}, 566 (1989).\\
S. Myers, A. Readhead and A. Lawrence, {\it Ap. J.} {\bf 405}, 8 (1993).
\bibitem{bcdes94} R. Bond, R. Crittenden, R. Davis, G. Efstathiou and P.
Steinhardt,\\Phys.
Rev. Lett. {\bf 72}, 13 (1994).
\bibitem{f71} W. Feller, `An Introduction to Probability Theory
and its Applications', New York: Willey, (1971).
\bibitem{gpjbbbs90} J.R.Gott III, C. Park, R. Juszkiewicz, W. Bies, D.
Bennett, F. Bouchet, A. Stebbins, {\it Ap. J.} {\bf 352}, 1 (1990).
\bibitem{mpb94}R. Moessner, L. Perivolaropoulos and R. Brandenberger,
Bulletin Board: astro-ph@babbage.sissa.it - 9310001,
{\it Ap. J.} {\bf 425}, 365 (1994).
\bibitem{gp94} A. Gilbert and L. Perivolaropoulos, CfA preprint, submitted to
{\it Astroparticle
Physics} (1994).
\bibitem{pfour} L. Perivolaropoulos,  {\it M.N.R.A.S.} {\bf 267}, 529 (1994).
\bibitem{pvari}
A. Gangui and L. Perivolaropoulos, CFA-3915,
Submitted to Astrophys. J.
Bulletin Board: astro-ph@babbage.sissa.it - 9408034, (1994).
\end{thebibliography}
\end{document}